\documentclass[%
reprint,
superscriptaddress,
 amsmath,amssymb,
 aps,
]{revtex4-2}
\usepackage[usenames,dvipsnames]{color}
\usepackage{graphicx}
\usepackage{dcolumn}
\usepackage{float}

\usepackage{booktabs}
\usepackage{makecell}
\usepackage{bm}
\usepackage{natbib}
\usepackage[utf8]{inputenc}
\usepackage{hyperref}

\begin{document}

\preprint{APS/123-QED}

\title{Discontinuous character of the ultrafast exciton Mott transition in monolayer WS$_2$}

\author{Subhadra Mohapatra}
\affiliation{Humboldt-Universität zu Berlin, Institut für Chemie, Berlin, Germany}

\author{Samuel Palato}
\affiliation{Humboldt-Universität zu Berlin, Institut für Chemie, Berlin, Germany}

\author{Nicholas Olsen}
\affiliation{Department of Chemistry, Columbia University, New York, USA}

\author{Julia Stähler}
\affiliation{Humboldt-Universität zu Berlin, Institut für Chemie, Berlin, Germany}
\affiliation{Fritz-Haber-Institut der Max-Planck-Gesellschaft, Abt. Physikalische Chemie, Berlin, Germany}
\author{Lukas Gierster}
\email{Corresponding author; lukas.gierster@hu-berlin.de}
\affiliation{Humboldt-Universität zu Berlin, Institut für Chemie, Berlin, Germany}


\date{\today}

\begin{abstract}
There are conflicting predictions and reports on the character of the exciton Mott transition (EMT) in monolayer transition metal dichalcogenides. It could be either a discontinuous or a continuous transition from the excitonic to the plasma phase, with important implications for devices such as photoswitches. To resolve the nature of the transition in monolayer WS$_2$, we study its ultrafast optical response upon resonant photoexcitation of the A exciton across a broad range of photoexcitation densities. In agreement with previously reported measurements we observe that the A exciton quenches gradually with increasing excitation density. However, a detailed lineshape analysis unveils an abrupt red shift in the transient peak positions of the A and B exciton resonances above an excitation density threshold. This is attributed to band gap renormalization arising from the formation of free charge carrier plasma, i.e., the EMT. The plasma phase decays with a time constant of 0.65~ps back into the excitonic state. The abrupt appearance of the plasma phase at the threshold density suggests that the EMT is a discontinuous and not a continuous transition. This work demonstrates how transient optical spectroscopy combined with lineshape analysis of two excitonic resonances simultaneously can be used to investigate the EMT in 2D materials.
\end{abstract}

\maketitle

\section{Introduction}
Strong light-matter interaction and large exciton binding energies make atomically thin TMDCs exciting systems for fundamental studies and applications \cite{ye2017recent, liu20192d, wu2015monolayer}. While at low charge carrier densities, the optical and electronic properties are governed by excitons, at a critical (Mott) density, enhanced screening weakens the Coulomb attraction between electrons and holes and leads to dissociation of the exciton gas into a metal-like electron–hole plasma - the
EMT. This transition has been intensely studied \cite{wang2019optical,li2021ultrafast,yu2020exciton, xu2025room, siday2022ultrafast,pekh2020electron, wang2019optical,bataller2019dense,sousa2023ultrafast,chernikov2015population,steinhoff2017exciton,guerci2019exciton}, both because device concepts such as excitonic optoelectronics, valleytronics, and nanolasers depend on the balance of free carriers and excitons \cite{handa2024spontaneous}, and because high densities enable exploration of many-body physics, including the electron–hole liquid with its exotic correlated properties \cite{yu2019room, arp2019electron}. The large binding energies in TMDCs yield strong excitonic signatures even at room temperature, suggesting that optical spectroscopy should readily probe the EMT. Yet for monolayers - with direct band gaps and the highest binding energies - it remains unclear whether the EMT evolves continuously with increasing excitation density, via coexistence of excitons and plasma, or discontinuously, as an abrupt or avalanche-like transition \cite{Rabinovich2024}. Resolving this is crucial for device design and for understanding photoexcited TMDC monolayers at high densities. Moreover, an ultrafast, discontinuous EMT could enable novel TMDC-based photoswitches and potentially access exotic quantum phases.

Theoretically, both continuous and discontinuous EMTs are possible. Material-realistic many-body calculations using GW-DFT and Bethe–Salpeter methods predict that a small exciton fraction (1~\%) dissociates below the Mott density, followed by an abrupt jump to full ionization \cite{steinhoff2017exciton}. Such an abrupt jump is calculated for WS$_2$, WSe$_2$, MoS$_2$ and MoSe$_2$ TMDC monolayers \cite{steinhoff2017exciton} and suggests a discontinuous character of the EMT. In contrast, a more conceptual study using a half-filled Hubbard model mapped to the exciton problem argues that the EMT character depends on the exciton binding energy \cite{guerci2019exciton}, which varies with TMDC composition. 

Experimentally, EMT studies have focused mainly on bilayers \cite{wang2019optical,li2021ultrafast,yu2020exciton, xu2025room, siday2022ultrafast}, heterostructures \cite{pekh2020electron, wang2019optical}, and bulk TMDCs \cite{dendzik2020observation, karmakar2021electron}. For monolayer TMDCs, however, only limited and conflicting observations exist. In MoS$_2$, continuous wave excitation reveals an abrupt switch from narrow exciton emission to broad electron–hole plasma emission at the Mott density \cite{bataller2019dense}, later linked to an electron–hole liquid \cite{yu2019room}. By contrast, WSe$_2$ and MoSe$_2$ monolayers show a continuous EMT, evidenced by continuously broadened time-integrated photoluminescence under ultrafast excitation \cite{sousa2023ultrafast}. Ultrafast time-resolved EMT measurements have so far been reported only for WS$_2$ \cite{chernikov2015population}, where pump–probe spectroscopy shows a gradual quenching of the A exciton resonance (AX) and the simultaneous emergence of stimulated emission below the optical gap due to plasma-induced band-gap renormalization (BGR) over a large ($>$1 order of magnitude) pump laser fluence range \cite{chernikov2015population}. The coexistence of exciton and plasma signals supports a continuous EMT in WS$_2$ \cite{guerci2019exciton}. These differing behaviors across monolayer compositions may reflect the Mott–Hubbard model \cite{guerci2019exciton}, but they contrast with \textit{ab initio} predictions of an abrupt EMT \cite{steinhoff2017exciton}.

The above observations do not permit firm conclusions. On the one hand, the time-resolved measurements in WS$_2$ show that the plasma decays on a few picosecond timescale, while the overall photoexcited state relaxes much more slowly \cite{chernikov2015population}. Thus, the continuous EMT inferred for WSe$_2$ and MoSe$_2$ from time-integrated PL must be interpreted cautiously: if the plasma rapidly relaxes into excitons, their luminescence will mix with plasma emission even if the initial transition is abrupt. 
On the other hand, in the time-resolved measurements on WS$_2$, the mere persistence of an excitonic resonance does not guarantee stable excitons, because collective Coulomb screening develops on ultrafast timescales comparable to the inverse plasma frequency \cite{huber2001many}. Transient excitons can even appear at metal surfaces \cite{schone2000time, cui2014transient}. In bulk WSe$_2$, above the EMT, time-resolved X-ray photoelectron spectroscopy reports exciton dissociation within about 100~fs \cite{dendzik2020observation}. Therefore, in TMDCs probed with femtosecond pulses, an exciton resonance may still appear-even in a pure plasma phase - albeit broadened, shifted, or weakened by photoexcited carriers.

Fortunately, due to the clearly defined exciton resonances in TMDCs, these changes can be tracked in the ultrafast time domain using line shape analysis, as previously demonstrated for low excitation densities below the EMT \cite{calati2023dynamic, ruppert2017role, cunningham2017photoinduced, sie2017observation}. Our previous work \cite{calati2023dynamic} showed that excitons, free charge carriers, and lattice heating can be disentangled in WS$_2$. For example, the presence of other excitons blue shifts the AX, while free charge carriers and lattice heating cause a red shift due to BGR. Applying such analysis to optical spectra across the EMT has so far been elusive. Furthermore, photoexciting one excitonic resonance can affect another excitonic resonance in the system. Transient absorption and 2D spectroscopy has shown that an A exciton population leads to an amplitude change of the second lowest energy B exciton resonance (BX) \cite{zou2023coherent, timmer2024ultrafast, lloyd2021sub, manca2017enabling}. The shift dynamics have not been analyzed yet. Interestingly, \textit{ab initio} calculations of optical pump-probe spectroscopy predict that the BX exhibits no shift upon photoexcitation of A excitons \cite{katsch2019theory}. This could potentially make the BX shift an exclusive spectator for BGR and hence, the existence of free charge carriers when tuning the photoexcitation density across the EMT.

Here, we report a systematic study of the EMT in monolayer WS$_2$ through fluence-dependent ultrafast transient absorption (TA) spectroscopy combined with a detailed line-shape analysis of both the AX and the BX. Using photoexcitation resonant with the AX, we photoexcite the system across a broad range of excitation densities, in which the EMT is expected to occur, and track the temporal evolution of the optical response in the ultrafast time domain. In the low laser fluence regime, the AX position and width is determined by the pump-induced A exciton population and the heating of the lattice, as observed before \cite{calati2023dynamic}. We confirm the theoretical prediction \cite{katsch2019theory} that the BX does not show an A exciton population-induced blue shift, but exclusively exhibits a red shift due to the BGR induced by lattice heating and free charge carriers, which can be well separated due to their different decay times. As the excitation density increases, a gradual reduction of the oscillator strength of the AX is observed, in agreement with previous measurements \cite{chernikov2015population}. This is attributed to population-induced bleaching. Far before the exciton resonance is completely bleached, the lineshape analysis of the differential spectra reveal an abrupt red shift of several 10's of meV of the AX and the BX at a critical excitation density of $n_\text{C}$~=~29~$\pm$~2~x~10$^{12}$~cm$^{-2}$ due to BGR upon plasma formation. The plasma decays within 0.65~ps, leading to a recovery of the excitonic phase. The BX, being exclusively sensitive to BGR, can be used to track the plasma population across all excitation densities. Besides the abrupt jump at the critical density, a small amount of plasma is detected already at low excitation densities. The trend closely follows the prediction by \textit{ab initio} many-body theory of Ref.~\cite{steinhoff2017exciton}. The abrupt change of the excitonic resonances suggests that the EMT in WS$_2$ has a discontinuous character. This, combined with the previously reported discontinuous EMT in MoS$_2$, suggest that the different exciton binding in these systems does not play a significant role in the EMT character, contrary to theoretical predictions \cite{guerci2019exciton}. The excitonic resonances persist across the Mott transition, which is consistent with a finite dissociation time of photoexcited excitons to plasma. Our data shows that the dissociation is completed within less than 200~fs. This work shows how optical spectroscopy can be used to investigate the Mott transition in 2D materials. In particular, we show that the inspection of several excitonic resonances at once can yield a comprehensive picture of the charge carrier dynamics across all excitation densities.
 
\section{Experimental Section}
To resolve the optical response of the WS\textsubscript{2} monolayer following resonant photoexcitation, we use TA spectroscopy, as sketched in Fig.~\ref{Fig:1}(a). TA spectroscopy is a pump-probe technique, where a first femtosecond laser pulse ("pump") photoexcites the system, and second laser pulse ("probe") monitors the photoinduced changes of the absorption at variable delays. The monolayer sample is photoexcited resonant with the A exciton, using a photon energy of 1.985~eV. A low fluence white-light continuum (WLC) probe pulse ranging from 1.8 to 4.5~eV is used to measure the absorption $A$ of the sample across multiple excitonic transitions. The pump-induced change of the absorption is recorded as $\Delta A=A_\text{pumped}-A_\text{unpumped}$.

The experimental setup was described in detail previously in Refs.~\cite{dobryakov2010femtosecond, kovalenko1999femtosecond}. Briefly, the laser source for TA is a regenerative amplified laser system (Astrella by Coherent), which outputs 30\ fs laser pulses at 800 nm wavelength with a repetition rate of 5\nolinebreak\ kHz. The fundamental beam is split to produce separate pump and probe beam paths. In the pump beam path, the desired wavelength is generated using an Optical Parametric Amplifier (Light Conversion TOPAS). In the probe beam path, the 800~nm light is frequency doubled and generates the WLC in a CaF\textsubscript{2} crystal. A delay stage (Physik Instrumente M-531.5IM) controls the time delay between the pump and probe pulse. The chirp of the white light pulses is corrected \textit{a posteriori} and computationally with about 20~fs precision as described in Ref.~\cite{kovalenko1999femtosecond}. The cross-correlation of pump and probe laser pulses is on the order of 50\nolinebreak\ fs (FWHM). 

To calculate the laser fluence the spot sizes of pump- and probe laser pulses are characterized at the sample position with a CCD camera. Pump- and probe pulses have a similar spot size of $r$=150\nolinebreak\ $\mu$m (D4$\sigma$). The laser fluence is calculated using 
\begin{equation}
F=\frac{2P}{AR}
\end{equation}
with $P$ the laser power, $A=2\pi r^2$ and $R$ the laser repetition rate. With the factor of 2 this formula gives the fluence at the center of a Gaussian beam profile (peak fluence). Since pump and probe spot sizes have approximately the same size in our experimental configuration, the edge of the probe beam samples areas with lower excitation density as well. Therefore, our fluence calculation (peak fluence) slightly overestimates the fluence. With this non-ideal setting our analysis probably overestimates the width changes and underestimates the peak shifts. However, as discussed in section \ref{sec:excitonic_regime}, the quantitative results are in agreement with previous studies in the excitonic regime. Moreover, apparently the setting does not affect the spectra so much as to inhibit a sharp Mott transition as shown in section \ref{sec:EMT} and is therefore also not expected to severely affect the Mott density determined in this work. 

The high-quality WS\textsubscript{2} monolayer on a fused silica substrate is prepared from its bulk counterpart using the gold-exfoliation technique, detailed in Refs.~\cite{desai2016gold, liu2019direct,olsen2025macroscopic}. This method provides high quality monolayers with less defects than commercially grown monolayers 
\cite{handa2024spontaneous}. The flake size is several hundred micrometers. All measurements are performed at room temperature ($ 22~^\circ \text{C}$).

\section{Results}
The results section is structured as follows. First, we present a data set to define the observables and determine the excitation density. Then we present the kinetics below and across the Mott density. 
\subsection{Data and excitation density}
\label{sec:data_and_excitation_density}
\begin{figure}[h]
\includegraphics[width=8cm]{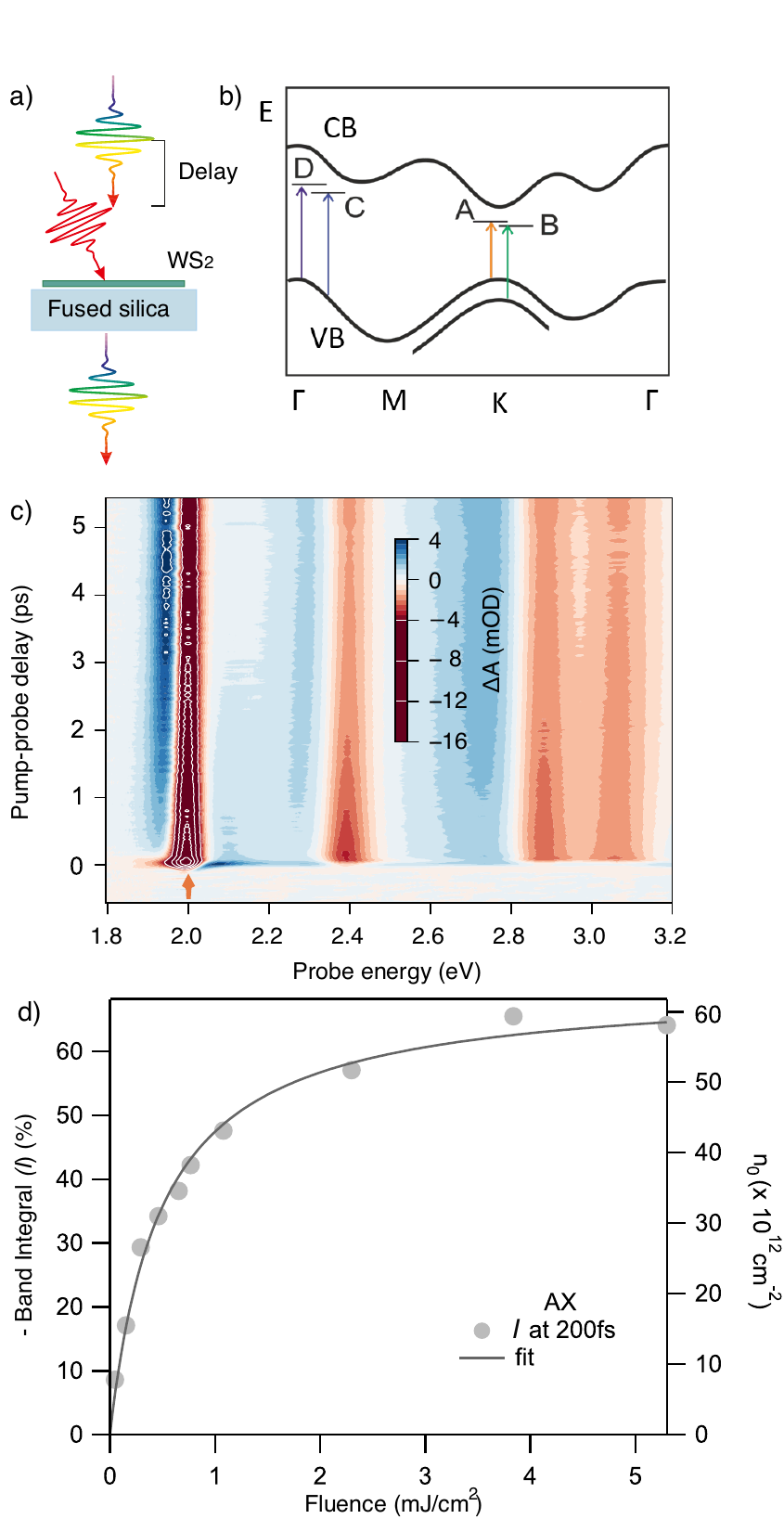}
\caption{\label{fig:epsart}\small 
(a) Sketch of TA applied to a WS\textsubscript{2} monolayer on a fused silica substrate.
(b) Band diagram and excitonic resonances (AX, BX, CX, DX) in the monolayer inserted into the one-particle band structure as photoionization energy levels of the excitons. (c) Energy and delay-dependent differential absorption spectra (color plot) of the monolayer photoexcited resonantly to the AX. (d) AX band integral and saturable absorber model to determine the excitation density.}
\label{Fig:1}
\end{figure}

An exemplary data set obtained for a single pump fluence at 150~$\mu J$/cm\textsuperscript{2} is shown in Fig.~\ref{Fig:1}(c). The differential absorption ($\Delta A$) is displayed as a function of pump-probe delay and probe energy. The false color plot shows weak positive (blue), and prominent negative (red) signals. They appear immediately at time zero (i.e., at the pump-probe overlap) and decrease partially within the first few picoseconds. A part of the signal surpasses the shown delay window and survives up to the longest measured delay of 60~ps (not shown). The negative signals are centered at around 1.99~eV, 2.38~eV, 2.88~eV, and 3.05~eV and align with the A, B, C, and D excitonic resonances, respectively, which can be observed in the steady-state absorption spectra of the WS\textsubscript{2} monolayer \cite{cunningham2017photoinduced, wang2018colloquium, calati2023dynamic} (cf. supplementary information section S1). Fig.~\ref{Fig:1}(b) shows the origin of these features. The AX and the BX are located in the K valleys in the band structure of the WS\textsubscript{2} monolayer. The energy difference between the two excitons is primarily attributed to spin-orbit coupling, which splits the valence band (VB) into two sub-bands. Besides the band edge excitons, C and D excitons are higher-lying optical transitions that form a bound state at the band nesting region along the $\Gamma$- K point and $\Gamma$ point, respectively \cite{wang2018colloquium}. Thus, our probe can observe several excitonic resonances simultaneously. In this paper, we focus on the AX and the BX; the BX will be shown to be a convenient observer of the EMT.

Before analyzing the charge-carrier dynamics that induce the changes to the optical absorption, it is necessary to determine the initial excitation density $n_0$ in the system for each pump laser fluence. $n_0$ is determined by how many photons are absorbed in the monolayer. In the linear limit (i.e., at low laser fluences) the absorptance at the pump wavelength can be determined by steady state spectroscopy \cite{calati2023dynamic}. However, strong photoexcitation can lead to changes of the absorption, as we have just demonstrated in the above exemplary dataset (see Fig.~\ref{Fig:1}(c). 
To quantify the pump-induced absorption change at the pump wavelength, we integrate the differential absorption signal around the AX resonance (1.8~to 2.1~eV) computing a band integral \cite{tanda2024evidence} and inspect it as a function of pump laser fluence. Here and in the rest of the paper we restrict our analysis to delays larger than 200~fs to avoid artifacts resulting from the coherences during the pump-probe overlap. Fig.~\ref{Fig:1}(d) shows the band integral of the AX at 200~fs relative to its steady state value. The absorption is reduced (note the flipped sign as indicated in the axis label) increasingly so at the higher pump fluence. 
The bleach grows non-linearly, approaching 60\% at the highest pump laser fluence of 5~mJ/cm$^2$. The solid line is a fit with the saturable absorber model 
    \begin{equation}
I(F)=\alpha\frac{F}{F+F_\text{sat}}
\label{sat_absorber}
\end{equation}
where $\alpha$ is the saturation amplitude. $F_\text{sat}$ is the characteristic fluence at which the absorption reaches half of its saturated value. 
The fit clearly describes the AX band integral trend very well.

The resolved trend can now be used to calculate the excitation density $n_0$. The band integral $I$ is assumed to be proportional to the excitation density $n_0$, due to ground state bleach as discussed below. The proportionality constant between $I$ and $n_0$ can be determined by comparing the low fluence limit of Eq.~(\ref{sat_absorber}) to the value determined from the steady-state absorptance. 

The resulting excitation density as a function of laser fluence is given on the right axis in Fig.~\ref{Fig:1}(d). The excitation density range of 7.75 to 43~x~10$^{12}$~cm$^{-2}$ covers the theoretically calculated Mott density of 8~x~10$^{12}$~cm$^{-2}$ for WS$_2$ on SiO$_2$ \cite{steinhoff2017exciton}.

The gradual increase of the bleach of the excitonic absorption is in agreement with previous measurements of Ref.~\cite{chernikov2015population}. Physically, this change could be due to i) ground state bleach, ii) due to a change of screening in the system (in particular upon undergoing the EMT), as well as iii) by a pump-induced temperature change. The latter is a weak contribution as we show below by analyzing the kinetics of the AX band integral. The fact that the saturable absorber model fits so well to the data strongly suggests that a simple population-induced bleach is at the origin of the oscillator strength reduction. Indeed, it can be noted that up to 100\% bleach are reached at high fluence and close to time zero (cf. supplementary information section S5), as expected for a few-level system at population saturation. Screening, instead, does seemingly not contribute to the AX bleach, as will be discussed in more detail at the end of the paper after having presented the discontinuous EMT. 

\subsection{Lineshape analysis and kinetics}
After introduction of the observables and the determination of the excitation density, we now move on to analyze the kinetics. Changes of the optical absorption of TMDC monolayers not only result from a change of the oscillator strength of excitonic resonances, but to a great extent by shift and broadening of these resonances \cite{calati2023dynamic, ruppert2017role, li2018two}. 
Access to these observables requires an approach beyond spectral integration, namely lineshape analysis. In the following we introduce our lineshape analysis model and apply it to the differential absorption data. We then proceed to analyze the kinetics of the pump-induced changes. To disentangle effects that simply scale with the number of excited quasiparticles from those that depend on their character (excitonic vs free carrier) as well as the pump-induced lattice heating, we first determine the quasiparticle population lifetime. Finally we examine the excitonic regime and the changes of the kinetics across the Mott density. 
\subsubsection{Model for lineshape analysis}
\begin{figure}
    \centering
    \includegraphics[width = 8 cm]{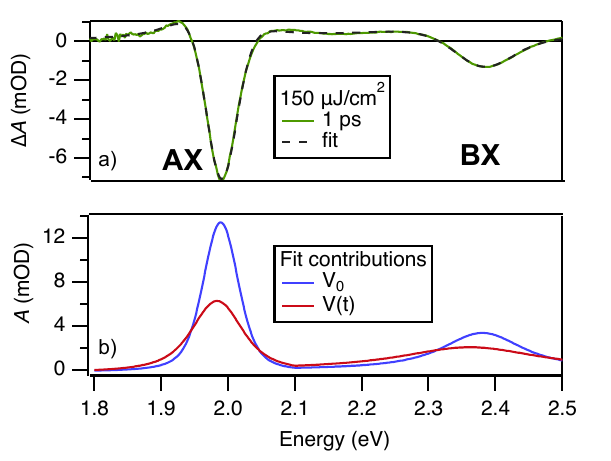}
    \caption{
    (a) Differential absorption spectrum, at a delay of 1~ps and pump fluence of 150~$\mu J$/cm$^2$ ($n_0$=15.5~x~10$^{12}$~cm$^{-2}$), fitted with the difference of two Voigt functions $V(t)-V_0$ (according to Eq.~\ref{equation1}). (b) $V(t)$ and $V_0$ contribution of the fit.}
    \label{fig:model}
\end{figure}

Fig.~\ref{fig:model}(a) shows an exemplary differential absorption spectrum of the AX and the BX at a pump-probe delay of 1~ps, extracted from the two-dimensional dataset in Fig.~\ref{Fig:1}(c). The shape is characterized by strong negative signals at the resonances, but also contains positive signals, which are clearly visible at the low energy side of the AX and in between the AX and the BX. In the easiest attempt to model the data, we assume that the differential spectrum is solely due to a pump-induced modification of the ground state exciton resonances. The steady-state resonances are described by Voigt profiles (cf. supplementary information section S1). We therefore model the differential spectra as the difference between Voigt profiles
\begin{equation}
\Delta A = V(t)-V_0. 
\label{equation1}
\end{equation}
Note $V_0$ is not equal to the steady state response $V_s$ as discussed below. $V(t)$ and $V_0$ are composed of the sum of the Voigt profiles of the AX and the BX:
\begin{equation}
        V= V_{A}(S_A, \gamma_A, \sigma_A, E_A)+ V_{B}(S_B, \gamma_B,\sigma_B, E_B)
\end{equation}
Here \emph{S}\textsubscript{A/B} is the oscillator strength, $\gamma_{\text{A/B}}$ the respective Lorentzian width, $\sigma_{A/B}$ the Gaussian width and $E_{\text{A/B}}$ are the AX and BX peak positions. For $V(t)$ the parameters are dependent on the pump-probe delay $t$ and for $V_0$ they are kept the same for all delays. It should be noted that the Voigt profile is the convolution of a Gaussian and a Lorentzian profile and therefore has two width parameters. Since the difference between the two widths is very subtle and prevents convergence, a common and time-independent Gaussian width ($\sigma_{A/B}$=$\sigma$) for both AX and BX is used while varying the Lorentzian width ($\gamma_{\text{A/B}}$). This is sufficient to describe the differential absorption data. 

Applying this model to fit the exemplary differential spectrum in Fig.~\ref{fig:model}(a) leads to a satisfactory match with the data. The corresponding $V(t)$ and $V_0$ are shown in the Fig.~\ref{fig:model}(b). Before interpreting the results, i.e., the extracted non-equilibrium response $V(t)$, it is important to understand the $V_0$ contribution. $V_0$ does not vary as a function of pump-probe delay, but is extracted for each pump fluence using a global fitting procedure,
in which several spectra between 200~fs to 1~ps 
are fitted using Eq.~\ref{equation1} with a common $V_{0}$ and a time varying $V(t)$. The position and width of the global fit-extracted $V_0$, although showing only a small variation with pump fluence, agree closely with those of the steady state absorption (cf. supplementary information section S2). 
However, its oscillator strength increases monotonically with pump-laser fluence (cf. supplementary information section S2), and approaches the absorption in the steady-state, but always remains below it. This is reasonable if one considers that a great part of the sample remains unexcited at low fluences. In the differential absorption signal $\Delta A$ as defined in the experimental section, the parts of the optical response cancel out if they are not modulated by the pump laser beam. Therefore, at low pump fluences, the oscillator strength of $V_0$ is expected to be lower than the steady-state value (i.e. the oscillator strength of $V_s$). As the pump laser fluence increases, so does the excitation density, until almost every probe photon encounters a spot that is affected by photoexcitation with the pump laser beam. This photoselection effect is reflected in the growth of $V_0$ with pump laser fluence. We expect, and observe that the $V_0$ oscillator strength asymptotically approaches the steady-state value. We emphasize that replacing the $V_{0}$ with the known steady state spectra of the monolayer using their full oscillator strength doesn't fit the differential data well. In contrast, the description of the differential data with the $V_0$ determined by the global fit procedure, is excellent across a broad range of delays and fluence (cf. supplementary information section S3), strongly supporting this way of analyzing the data. 

By fitting $V(t)$ at all pump-probe delays with a common $V_0$, the following transient parameters are extracted:
\begin{eqnarray*}
    \Delta E_{A,B} (t) = E_{A,B} (t)-E_{A,B}^0\\
    \Delta \gamma_{A,B}(t) = \gamma_{A,B}(t) - \gamma_{A,B}^0\\
    \Delta S_{A,B} = (S_{A,B} - S_{A,B}^0)/S_{A,B}^{0}
    \label{eqn:para}
\end{eqnarray*}
where $\Delta E_{A,B}$, $\Delta \gamma_{A,B}$ and $\Delta S_{A,B}$ correspond to the transient shift, broadening and oscillator strength change of each excitonic resonance (A, B), respectively. 
For the exemplary fit in Fig.~\ref{fig:model}(b), the fit retrieved shift, broadening and changes in oscillator strength magnitudes are tabulated in Table~\ref{tab: table1}. $V(t)$ is slightly red shifted with respect to $V_0$, substantially broadened, and, in the case of the AX, it also has a lower oscillator strength. The oscillator strength of BX is slightly increased, opposite to that of AX and contrary to the initial expectations from inspecting the data.
\begin{table}[]
\centering
\begin{tabular}{c|c|c|c} 
\toprule
\makecell[c]{Excitonic\\Resonance} & $\Delta E_{A,B}$ (meV) & $\Delta\gamma_{A,B}$ (meV) & $\Delta S_{A,B}$ (\%)\\
\midrule
AX &  -5.4 $\pm$ 0.1 & 39.5 $\pm$ 0.2 & -21 $\pm$ 0.3\\
BX &  -18 $\pm$  0.9 & 119.0 $\pm$ 1.5 &12.8 $\pm$ 0.8\\
\bottomrule
\end{tabular}
    \caption{Fit-retrieved transient shift, broadening and oscillator strength change of AX and BX at a pump-probe delay of 1 ps and a pump fluence of 150~$\mu J$/cm$^2$ ($n_0$=15.5~x~10$^{12}$~cm$^{-2}$).}
    \label{tab: table1}
\end{table}
We note in passing that this calls into question previous interpretations of absorption data, where the reduction of signal at the BX energy is attributed to a reduction of its oscillator strength, due to - for example - an AX to BX population transfer \cite{guo2019exchange,  timmer2024ultrafast, lloyd2021sub, manca2017enabling}. Discussing the BX oscillator strength in greater detail, however, is beyond the scope of this article. Instead, we now apply the model to explore the population lifetime of A excitons, and thereafter the shift and broadening kinetics in the excitonic regime and across the Mott transition. 
\subsubsection{Population lifetime}
\label{sec:population_lifetime}
In section \ref{sec:data_and_excitation_density}, we have established that the AX oscillator strength reduction is due to ground state bleach. We now examine the quasiparticle population lifetime by inspecting the recovery dynamics. We use the oscillator strength determined with the lineshape analysis. Almost identical results are achieved from analyzing the AX band integral kinetics (not shown). 

\begin{figure}[h]
\centering
\includegraphics[width= 8 cm]{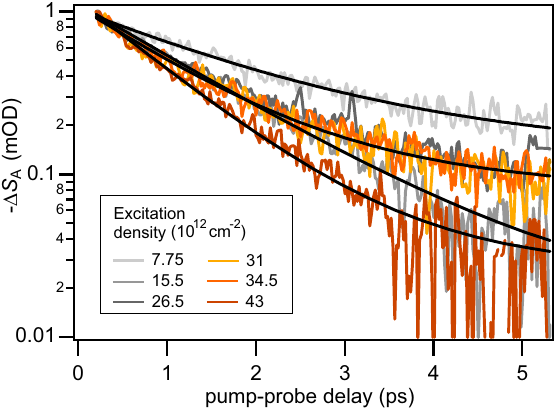}
\caption{
Ultrafast relaxation of the AX oscillator strength change, i.e., the ground state bleach, for different initial excitation densities. To evaluate the relaxation time, the traces are normalized at 200~fs and fitted with a with a single decay exponential function (see text). For clarity, the fit functions are shown only for four traces, as black lines.}
\label{fig:Figure_GSB}
\end{figure}

Fig.~\ref{fig:Figure_GSB} shows the time-dependent oscillator strength change of the AX for different initial excitation densities. The intensity transients are normalized at 200~fs and displayed on a logarithmic scale. The oscillator strength change recovers to about 10~\% within the first 5~ps for all excitation densities - slightly slower for the lowest and slightly faster for the highest excitation density. The remaining 10~\% decay on a 10's of~ps timescale (not shown) and are attributed to pump-induced lattice heating, which cools on a 20-30~ps time scale, as established in the subsequent section. 

To extract the population lifetime, we 
fit the data using a single exponential decay function:
\begin{equation}
    y(t)=A_\text{pop}\ \text{exp}(-t/\tau_\text{pop}) + y_0
\label{eq:population_decay}
\end{equation}
with the population decay time constant ($\tau_\text{pop}$) and an offset $y_0$ accounting for the remaining oscillator strength bleach. The fit describes the data well. At an excitation density of 15.5 x 10\textsuperscript{12} cm\textsuperscript{-2}, a dataset which we analyze in detail in the next section, the extracted population lifetime is $\tau_\text{pop}=1.4\pm0.2$~ps. For the highest excitation density the lifetime reduces to $\tau_\text{pop}=1.1\pm0.2$~ps.
These values are in the range of those determined with time-resolved photoelectron spectroscopy measurements at room temperature \cite{bange2023ultrafast} as well as other optical studies at comparable excitation densities \cite{wang2024ultrafast}.  From the above, we have determined the population lifetime. This enables us to interpret the shift and broadening kinetics in the following.

\subsubsection{Excitonic regime}
\label{sec:excitonic_regime}
\begin{figure*}
\includegraphics[width=17 cm]{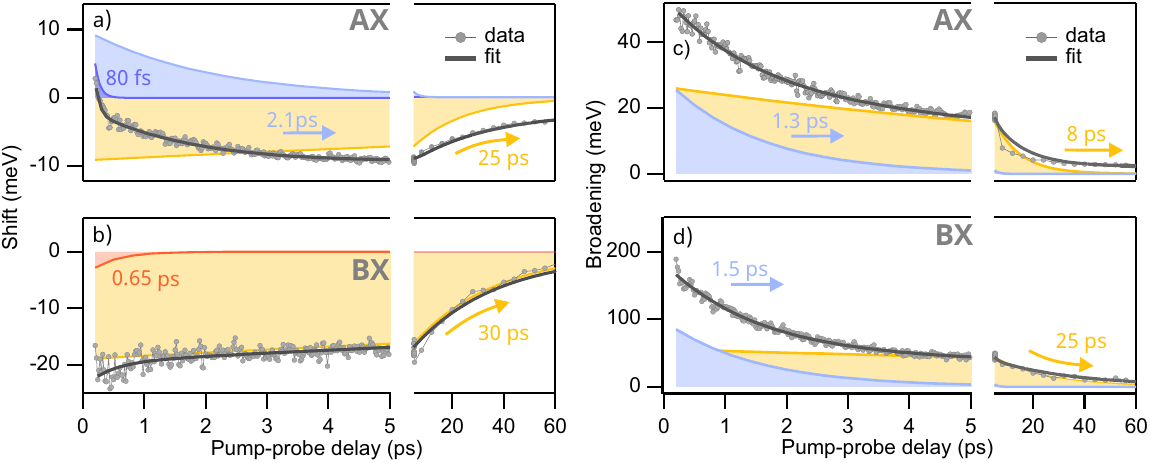}
\caption{\label{fig:epsart}Shift (a,b) and broadening (c,d) of A and B exciton resonances at a low excitation density (15.5~x~10\textsuperscript{12}~cm\textsuperscript{-2}) as a function of pump-probe delay. The data is fitted with a sum of exponential decay functions as described in the text. The individual components of the fit are illustrated with the shaded areas.}
\label{fig:excitonic_regime}
\end{figure*}
We start by characterising the optical response in the excitonic regime 
before proceeding to high excitation densities at which the Mott transition occurs. We inspect the peak shifts and width changes of the AX and the BX as a function of pump-probe delay for a low excitation density of 15.5~x~10$^{12}$~cm$^{-2}$. As outlined in the introduction, based on previous experimental findings \cite{calati2023dynamic, ruppert2017role} and theoretical predictions \cite{katsch2019theory}, we expect a blue shift of the AX in the presence of A excitons - but not of the BX. Furthermore, we expect a red shift of both the AX and the BX due to the pump-induced lattice heating, which causes BGR. In addition, both population and pump-induced lattice heating are expected to lead to broadening \cite{calati2023dynamic, selig2016excitonic, sim2013exciton, gupta2019fundamental, moody2015intrinsic}. In a collisional broadening picture, this should affect both excitonic resonances simultaneously. It should be noted that excitons and phonons (as well as free carriers which will be formed at higher excitation densities) are individual sets of quasiparticles, each evolving with their specific time constants. The observables, i.e., the optical resonances, see the superposition of the effects of all quasiparticles. We will show in this section empirically that all kinetics can be described with the superposition of exponential decays, each accounting for the temporal evolution of one set of quasiparticles. 
Based on this we will develop a global fit function in the next section that works across all excitation densities and delays. This will allow us to disentangle the contributions of excitons, free carriers and lattice heating to the optical response across the Mott density. 

Fig.~\ref{fig:excitonic_regime}(a) shows the overall shift of the AX resonance. Prior to going through the description of the dynamics in detail, let us note that we refer to shift components with a 
positive amplitude as “blue shifts” and those with a 
negative amplitude as “red shifts.” Because the measured dynamics arise from the superposition of multiple components with different decay constants, the overall shift at a given pump–probe delay may be either red or blue. In particular, a rapidly decaying initial blue shift superimposed on a more slowly decaying red shift can produce the appearance of a rising red shift (and \emph{vice versa}).


The overall AX shift firstly (at 200~fs) exhibits a few meV blue shift ($\Delta E_A>0$), which decays quickly. This is followed by the apparent rise of a red shift on a few-ps timescale and, finally, equilibration dynamics on a timescale of 10's of~ps. 
We find that the observed dynamics of the AX shift can be best fitted with a sum of three exponential decay functions, with two blue shift components and one red shift component, plus an offset. 
In preparation for the global fit developed later, we write:
\begin{equation}
    \Delta E_\text{A}(t) = B_0\ e^{-t/\tau_0}+B_1\ e^{-t/\tau_1} + R\ e^{-t/\tau_2}+y_0
    \label{equation: AX_shift}
\end{equation}
The resulting fit is shown as black solid line on top of the data in Fig.~\ref{fig:excitonic_regime}(a). It describes the data well. The figure also contains a decomposition of the total fit function into the contribution of each exponential to the fit (blue and orange curves). For the AX shift, the main part of the observed dynamics can be described with a blue shift decaying with $\tau_1=2.1\pm 0.1$~ps plus a red shift decaying with $\tau_2=25\pm 1 $~ps. In addition there is a very small contribution of a blue shift at early time delays decaying with $\tau_0=80\pm 10\ \text{fs}$ as well as a longer-lived signal, captured by $y_0$. $\tau_1$ is close to the population lifetime $\tau_\text{pop}=1.4\pm 0.2\ \text{ps}$ established from the ground state bleach recovery in the preceding section. $\tau_2$ is in agreement with the cooling time constant $\tau_\text{cool}$ of several tens of picoseconds resolved in previous work \cite{calati2023dynamic,ruppert2017role}. These two contributions are therefore attributed to an A exciton population-induced blue shift and the lattice heating-induced BGR, respectively. The lattice heating-induced red shift amplitude $R$ is determined to be -18.3$~\pm~0.1$ meV, which is in agreement with the value reported in Ref.~\cite{ruppert2017role} at a similar pump fluence. Using a proportionality constant derived from equilibrium measurements, see Ref.~\cite{ruppert2017role} and references therein, we can estimate that the pump excitation at this fluence leads to an increase in lattice temperature by 30~K at time zero. Beyond these two contributions, the extracted decay time constant of the small initial blue shift is in agreement with the scattering time constant of K-valley electrons into the $\Sigma$ Valley at the center of the Brillouin zone \cite{madeo2020directly, wallauer2021momentum}. 
We refer to it as $\tau_\Sigma$ in the following. The remaining signal after 60\ ps, captured by $y_0$ with the fit is commensurate with lifetimes of defect trapped carriers, which range up to ns \cite{wang2016radiative, palummo2015exciton}. As the $\tau_\Sigma$ blue shift component and the remaining signal due to defects are essentially out of the range of recorded delays, we do not discuss them further. Overall, as expected for the excitonic regime, the decay dynamics are dominated by the pump-induced excitons and lattice heating, each following first order kinetics. The two contributions can be well separated, once due to the different timescale, but also because they induce a shift of the optical resonance that is of opposite sign: In the main section of the accessed delay window, the AX shift dynamics are the superposition of a blue shift due to the A exciton population decaying with $\tau_\text{pop}$ and a red shift due to pump-induced lattice heating decaying with $\tau_\text{cool}$ on a tens of picoseconds timescale.

We now turn to the BX shift. As shown in Fig.~\ref{fig:excitonic_regime}(b) it exhibits a red shift of $-$20\ meV at 200\ fs. Except for a very small part of the red shift that decays in the first picoseconds, relaxation takes on a timescale of 10's of ps. We find that the BX shift is best described with the sum of two exponential decay functions, with two red shift components. 
Again, as preparation for the global fit employed later, we write this down explicitly:
\begin{equation}
    \Delta E_\text{B}(t) = R_1\ e^{-t/\tau_1}+R_2\ e^{-t/\tau_2}
    \label{equation: BX_shift}
\end{equation}
The resulting fit is shown as black solid line on top of the data in Fig.~\ref{fig:excitonic_regime}(b). The decomposition of the fit function into the individual components (solid red and yellow lines with shaded area) reveals that the main part of the BX red shift relaxes with a time constant of $\tau_2=30\pm 1\ \text{ps}$ to the equilibrium, in agreement with $\tau_\text{cool}$, the lattice cooling time constant established above. 
A very small part of the shift ($R_1=-$2 meV) relaxes with a time constant of $\tau_1=0.65\pm 0.10\ \text{ps}$. This is much faster than the population relaxation time constant $\tau_\text{pop}=1.4 \pm 0.2\ \text{ps}$. 
As shown below for the region above the Mott limit, $\tau_1$ can be related to the decay of free carriers. A small population of free carriers has been detected before in the low excitation density regime and has been attributed to defect-ionized excitons \cite{handa2024spontaneous}. Remarkably, the BX exhibits no component decaying on the timescale of the population decay. This is contrary to the AX, but as theoretically predicted \cite{katsch2019theory}. It is the absence of BX shift dynamics on few picoseconds-timescales that allows the detection of a very small contribution of the free carrier population, which would likely be obscured otherwise.  

To further corroborate that the dynamics in the excitonic regime are dominated by the pump-induced excitons and lattice heating, we also inspect the broadening. The broadening for the AX and the BX is shown in Fig.~\ref{fig:excitonic_regime}(c,d). Both excitonic resonances are already broadened at 200\ fs and the broadening relaxes on a picosecond time scale. Fits with a biexponential decay are shown as black solid line for both panels. They clearly describe the data well. The resulting time constants are $\tau_1=1.3\pm 0.1\ \text{ps}$, $\tau_2=8\pm 1\ \text{ps}$ for the AX, and $\tau_1=1.5\pm 0.1\ \text{ps}$, $\tau_2=25\pm 1\ \text{ps}$ for the BX. These two time constants agree well with $\tau_\text{pop}$ and $\tau_\text{cool}$ established above. The finding that $\tau_1$ equals $\tau_\text{pop}$ implies a proportionality between broadening and exciton population, in agreement with previous reports \cite{calati2023dynamic, moody2015intrinsic, gupta2019fundamental, Erkensten2021exciton}. %

Taken together, shift and broadening kinetics in the excitonic regime can be majorly attributed to pump-induced A excitons and pump-induced lattice heating as observed in previous studies. The decay of the pump-induced changes can be described empirically with a sum of exponential decay functions, each accounting for different processes in the sample that superimpose to a total shift and broadening of the excitonic resonances, respectively. 
Notably, the BX shift is, on the one hand, insensitive to the A exciton population as predicted theoretically \cite{katsch2019theory}. On the other hand, it is sensitive to the changes of the underlying single-particle band structure by BGR induced by lattice heating and free carriers. These properties qualify the BX resonance to be a spectator of the Mott transition. 

\subsubsection{Exciton Mott transition}
\label{sec:EMT}
Upon enhancing the excitation density and crossing the Mott density, photoexcitation will initially still create A excitons, but they subsequently dissociate into free carriers, i.e., an electron-hole plasma. The electron-hole plasma induces BGR and is expected to red shift the excitonic resonances \cite{calati2023dynamic}. We restrain the discussion here to the AX and the BX shift. The broadening also shows strong changes at the same excitation density \cite{mohapatra2026Photoinduced} (cf. supplementary information section S7). A detailed discussion is out of the scope of the present publication. For the shift analysis presented in the following, it should be noted that the dissociation is expected to happen on a screening timescale of tens of femtoseconds as mentioned in the introduction \cite{huber2001many}. Since we only analyze pump-probe delays $>$200~fs, we expect to observe only the decay and not the formation of the plasma. 

\begin{figure}[h]
\includegraphics[width=8cm]{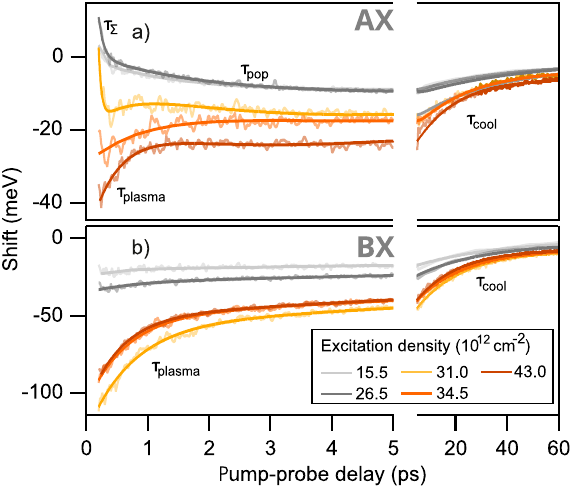}
\caption{\label{fig:wide} AX (a) and BX (b) shifts as a function of pump-probe delay for different initial excitation densities, together with fits (solid lines). The fit functions are the sum of four and two exponential decays for AX (see Eq.~\ref{eqn:AX_shift}) and BX (see Eq.~\ref{eqn:BX_shift}), respectively.}
\label{fig:Exciton_Mott_transition_1}
\end{figure}

Fig.~\ref{fig:Exciton_Mott_transition_1} shows the AX and the BX shift as a function of pump-probe delay for different excitation densities. The low excitation density trace at 15.5~x~10$^{12}$~cm$^{-2}$ discussed in the last section is shown in light grey as a reference. Roughly doubling the excitation density to 26.5~x~10$^{12}$~cm$^{-2}$ does not change the dynamics qualitatively, neither for the AX (Fig.~\ref{fig:Exciton_Mott_transition_1}(a)) nor the BX (Fig.~\ref{fig:Exciton_Mott_transition_1}(b)).

However, already at a slightly higher excitation density of 31~x~10$^{12}$~cm$^{-2}$, the dynamics change dramatically: the corresponding yellow trace shifts, down, up, down, and up again with various timescales. As will be clear later on, this is due to the superposition of multiple peak shifts caused by different physical phenomena. At an excitation density of 31~x~10$^{12}$~cm$^{-2}$, the shifts - coincidentally - have comparable amplitudes.
Even higher excitation densities result in a clear initial red shift of $-$30 to $-$40~meV, which partially decays within ca. 1~ps. After this initial relaxation, the dynamics show a similar decay as for low excitation densities.  

At the same critical excitation density of 31~x~10$^{12}$~cm$^{-2}$, the initial red shift of the BX jumps abruptly from $-$30~meV to about $-$100~meV. For even higher excitation densities the initial red shift does not increase further. The enhanced initial red shift decays with ca. 1~ps. After this initial relaxation, the dynamics also show a similar decay as for low excitation densities, 
i.e.,~a red shift that decays into equilibrium on a timescale of tens of picoseconds. 

Note that the abrupt red shift is also visible in the raw data: See supplementary information section S4 for a comparison of the spectra below and across the critical excitation density, together with a demonstration of how the lineshape analysis captures the spectral changes. 

The abrupt appearance of an enhanced red shift in both excitonic resonances, which quickly decays with a similar rate, strongly suggests that the excitation density has crossed the Mott limit. The pump laser beam now creates a sufficiently large excitation density, such that initially excited excitons dissociate into free carriers. This apparently happens within less than 200~fs and induces a red shift of both excitonic resonances due to BGR. The red shift decays within a picosecond and is followed by shift dynamics alike to those in the low excitation density regime. In particular, the AX shift also shows a weak few-ps blue shift contribution as $\tau_1$ in the previous section \ref{sec:excitonic_regime} that was attributed to the exciton population decay ($\tau_\text{pop}$ component, cf. grey traces and yellow trace in Fig. \ref{fig:Exciton_Mott_transition_1} (a)). This indicates the decay of the plasma and subsequent re-establishment of the excitonic phase.  

In the following, we develop a fit model that allows to describe the peak shifts quantitatively. It is based on three contributions, i.e. lattice heating, excitons and the plasma. We will prepare for a global fit across all excitation densities and delays to determine the time constant associated with the decay of the plasma phase and follow the evolution of the plasma as a function of excitation density. This will finally allow discussing the character of the transition. 

The basic assumption for the fit model is that the plasma population ($n_\text{plasma}$) either builds up an exciton population ($n_\text{AX}$) and then subsequently recombines, or directly decays into the ground state by recombination. This can be represented as the following coupled rate equations:
\begin{eqnarray}
\frac{dn_\text{plasma}}{dt}&&=-\frac{1}{\tau_\text{AX}} n_\text{plasma}- \frac{1}{\tau_\text{pop}}n_\text{plasma}\\
\frac{dn_\text{AX}}{dt}&&=\frac{1}{\tau_\text{AX}}n_\text{plasma} - \frac{1}{\tau_\text{pop}}n_\text{AX} ,
\end{eqnarray}
The excitons and free charge carriers are assumed to have the same recombination time constant $\tau_\text{pop}$, which is justified empirically as the ground state bleach recovery changes only slightly as a function of excitation density (cf. Fig. \ref{fig:Figure_GSB}). The solution of the rate equation system is
\begin{eqnarray}
n_\text{plasma} (t) &&=n_\text{plasma}^0\text{e}^{-\frac{t}{\tau_\text{plasma}}}\label{XXX1}\\
n_\text{AX} (t) &&= \left[n_\text{plasma}^0 (1- \text{e}^{-\frac{t}{\tau_\text{AX}}})+n_\text{AX}^0\right]\text{e}^{-\frac{t}{\tau_\text{pop}}}\label{XXX}
\end{eqnarray}
with
\begin{equation}
    \tau_\text{plasma}=\left(\frac{1}{\tau_\text{AX}}+\frac{1}{\tau_\text{pop}}\right)^{-1}
\end{equation}
$n_\text{plasma}^0$ is the initial plasma population directly after the dissociation process. $n_\text{AX}^0$ is the exciton population that remains from the dissociation process. This is necessary to cover the case of the low excitation density regime, where $n_\text{AX}^0$ dominates. Furthermore, we can cover the cases of a discontinuous or continuous Mott transition, where the pump-induced A excitons are dissociated completely at the Mott density or to a continuously increasing fraction upon increasing the excitation density, respectively. 

We now deduce fit functions for our observables, namely the AX and the BX shift, based on the rate equation model above and incorporating the additional contributions to the AX and BX shift derived with the empirical fits in the last section. We assume that the plasma population leads to BGR and hence a red shift of both the AX and the BX, proportional to $n_\text{plasma}$. Furthermore, the AX (but not the BX) exhibits a blue shift proportional to $n_\text{AX}$. As established in the last section, pump-induced lattice heating leads to an additional red shift contribution of both excitonic resonances, which decays exponentially in time with $\tau_\text{cool}$. With this, the fit function of the BX shift becomes a sum of two exponential decays:
\begin{eqnarray}
\Delta E_\text{B}(t) =&&{R}^\text{B}_\text{plasma} \text{e}^{-t/\tau_\text{plasma}}\notag\\
+&&{R}^\text{B}_\text{heat}\text{e}^{-t/\tau_\text{cool}}.
\label{eqn:BX_shift}
\end{eqnarray} 
Here ${R}^\text{B}_\text{plasma}$ and ${R}^\text{B}_\text{heat}$ are the plasma and lattice temperature-induced red shift amplitude, respectively. ${R}^\text{B}_\text{plasma}$ is proportional to the initial plasma population $n_\text{plasma}^0$ via Eq. \ref{XXX1}. This fit function is identical to the one derived empirically in the previous section (cf. Eq.~\ref{equation: BX_shift}) to describe the dynamics in the excitonic regime. 
This enables systematic tracking of the shift amplitudes over the entire excitation density range further below.

The fit function for the AX shift has three more components due to the sensitivity to the AX population $n_\text{AX}$, an additional fast ($\tau_\Sigma=80$~fs) decaying blue shift at early delays \footnote{It should be noted that above the critical excitation density of 31~x~10$^{12}$cm$^{-2}$, this component seems to be missing. This likely just because it is accelerated and does not appear at 200\ fs anymore (cf. supplementary information section S6)} and a long-lived component $y_0$, cf. the previous section. This leads to a sum of four exponential decays associated with $\tau_\text{plasma}$, $\tau_\text{pop}$, $\tau_\text{cool}$ and $\tau_\Sigma$, plus an offset:
\begin{eqnarray}
\Delta E_\text{A}\!&=&\!{R}^\text{A}_\text{plasma} \text{e}^{-t/\tau_\text{plasma}} \notag\\
&+&\!{B}_\text{pop}^\text{A}\text{e}^{-t/\tau_\text{pop}} \notag\\
&+&\!{R}^\text{A}_\text{heat}\text{e}^{-t/\tau_\text{cool}}+{B}^\text{A}_{\Sigma}\text{e}^{-t/\tau_{\Sigma}} + y_0.
\label{eqn:AX_shift}
\end{eqnarray}
Here, ${R}^\text{A}_\text{plasma}$ and $R^\text{A}_\text{heat}$ are the plasma and lattice temperature-induced red shift amplitude, respectively. As for the BX shift, ${R}^\text{A}_\text{plasma}$ is proportional to the initial plasma population $n^0_\text{plasma}$, which enters via Eq.~\ref{XXX1} and Eq.~\ref{XXX}. 
$B^\text{A}_\text{pop}$ is a blue shift caused by the A exciton population term Eq.~\ref{XXX}. Not trivially though, it is proportional to the sum of $n^0_\text{plasma}$ and $n^0_\text{AX}$, which is equal to the total pump-induced excitation density. 
This is because the model assumes that the plasma decays partially by forming A excitons. This exciton population then blue shifts the AX resonance the same way as initially created excitons. 
Compared to the empirical fit in the previous section, which perfectly described the dynamics in the excitonic regime, an additional exponential decay is introduced to describe the plasma-induced red shift ${R}^\text{A}_\text{plasma}$. This is forced to be zero at low excitation densities to avoid overfitting. 

We now fit the shifts for both the AX and the BX by optimizing $\tau_\text{plasma}$ and $\tau_\text{cool}$ globally across all excitation densities and together for the AX and the BX shift. Here, $\tau_\text{pop}$ is determined from the ground state bleach recovery of the AX (section \ref{sec:population_lifetime}). $\tau_\Sigma$ is fixed at 80 fs. The fits are shown with the solid lines in Fig.~\ref{fig:Exciton_Mott_transition_1}. Clearly, they describe the peak shifts for both excitonic resonances well. The retrieved plasma decay time is $\tau_\text{plasma}=0.65~\pm~0.1~\text{ps}$. The good match of the fit for both the AX and the BX shift using globally determined time constants shows that the peak shifts can be described with only three coupled contributions, i.e., the temporal evolution of lattice heating, excitons and the plasma. Particularly, the global fit shows that the initial red shift observed in both excitonic resonances at elevated excitation densities decays with the same rate. It therefore must be of the same physical origin, consistent with a plasma population causing BGR which affects both excitonic resonances. Notably, $\tau_\text{plasma}$ agrees well with the decay time of stimulated emission from plasma carriers in the renormalized conduction band, observed in optical spectroscopy at high excitation densities for the same sample system \cite{chernikov2015population}. 
\begin{figure}[h]
    \centering
\includegraphics[width=8 cm]{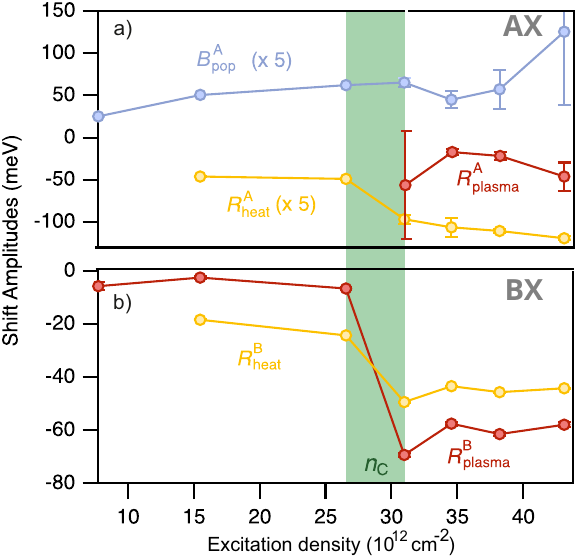}
    \caption{Shift amplitudes associated with the three decay time constants $\tau_\text{pop}$, $\tau_\text{cool}$ and $\tau_\text{plasma}$ (see text) for both AX (a) and BX (b) as a function of excitation density. $B^\text{A}_\text{pop}$ and $R^\text{A}_\text{heat}$ are multiplied by a factor of 5 to display the trend more clearly. The jump of all amplitudes except $B_\text{pop}^{\text{A}}$ across the Mott density ($n_\text{C}$) 
 indicates a discontinuous Mott transition.}
    \label{fig:Exciton_Mott_transition_2}
\end{figure}

Having established that the excitonic resonances detect a plasma and its decay at elevated excitation densities, 
we now tackle the question whether the dissociation ratio, and hence the plasma population, increases discontinuously or continuously upon increasing the excitation density. Clearly, in the time domain, a continuous decay of the plasma population 
is observed, leading to a gradual relaxation of the plasma-induced red shift for both the AX and the BX shift with $\tau_\text{plasma}$ and the re-establishment of the excitonic regime. This behaviour is independent on the character of the phase transition, because the plasma population decay is based on probabilistic events and not an abrupt coherent process such as optical excitation. 
To investigate the pump-induced effects extrapolated to the moment immediately after the dissociation, we inspect the relevant shift amplitudes as a function of initial excitation density. These are shown in Fig.~\ref{fig:Exciton_Mott_transition_2} for the AX and the BX, respectively. We can distinguish the following trends
\begin{itemize}
\item  The population-induced blue shift amplitude of the AX $B_\text{pop}^\text{A}$ (blue) increases as a function of excitation density. 
\item The plasma-induced red shift amplitude of the AX $R^\text{A}_\text{plasma}$, (red, top) which can only be resolved above 31~x~10$^{12}$~cm$^{-2}$, is at about -30 to -50 meV above this excitation density.
\item The plasma-induced red shift amplitude of the BX $R^\text{B}_\text{plasma}$ (red, bottom) is very small but finite below 31~x~10$^{12}$~cm$^{-2}$ and then jumps to $-$70~meV. It stays almost constant thereafter.
\item The lattice-heating induced red shift amplitudes of the AX and BX, i.e., $R^\text{A}_\text{heat}$ and $R^\text{B}_\text{heat}$ (orange), increase as a function of excitation density, but exhibit a distinct jump between 26.5~x~10$^{12}$~cm$^{-2}$ and 31~x~10$^{12}$~cm$^{-2}$.
\end{itemize}
All shift components besides ${B}^\text{A}_\text{pop}$ exhibit an abrupt change upon increasing the excitation density from 26.5~x~10$^{12}$~cm$^{-2}$ to 31~x~10$^{12}$~cm$^{-2}$. We, thus, define a critical density $n_\text{C}$~=~29~$\pm$~2~x~10$^{12}$~cm$^{-2}$ (green in Fig.~\ref{fig:Exciton_Mott_transition_2}). 
At $n_\text{C}$, a fundamental change in the kinetics occurs, showing that it separates two distinct phases of the system. In particular, the simultaneous and abrupt increase of the plasma-induced AX and BX red shift amplitudes ${R}^\text{A}_\text{plasma}$ and ${R}^\text{B}_\text{plasma}$ is consistent with a Mott transition occurring at $n_\text{C}$. 
Whether or not this transition is of continuous or discontinuous character is discussed in the following.

As outlined above, the red shift amplitudes ${R}^\text{A}_\text{plasma}$ and ${R}^\text{B}_\text{plasma}$ are proportional to the plasma population after the initial dissociation process of pump-induced A excitons. 
In the low excitation density regime below $n_\text{C}$, only few excitons dissociate, for example due to defect-induced ionization as already mentioned in the preceding section. The plasma-induced red shift amplitudes therefore are expected to be small. This matches well to the observation that the AX red shift ${R}^\text{A}_\text{plasma}$ can only be resolved above $n_\text{C}$. Due to the overlapping blue shift of the AX due to the A exciton population (which likely also leads to the high error bars in the ${R}^\text{A}_\text{plasma}$ and ${B}^\text{A}_\text{pop}$ amplitudes above $n_\text{C}$), a small plasma-induced red shift evolving at a similar time scale as the overall quasiparticle population decay with $\tau_\text{pop}$ is invisible. Notably, the BX red shift amplitude ${R}^\text{B}_\text{plasma}$ remains small but finite for all excitation densities below $n_\text{C}$. This observation shows, consistent with the individual fits in the preceding section, that in the excitonic regime a small plasma population is formed. The BX shift serves as an exclusive spectator of BGR and therefore allows to resolve the plasma-induced red shift across all excitation densities.

At the Mott density, exciton dissociation becomes favourable. The dissociation ratio, i.e. the plasma population compared to the total excitation density is expected to change. Crucially, the BX red shift amplitude ${R}^\text{B}_\text{plasma}$ at $n_\text{C}$ is abrupt, reminiscent of a step function as a function of the excitation density. This indicates an equally abrupt change of the dissociation ratio at $n_\text{C}$. This observation strongly suggest that the Mott transition has a discontinuous character and not a continuous one, in which case we would expect that ${R}^\text{B}_\text{plasma}$ changes gradually as a function of the excitation density. The sharp character of the transition is underscored by the abrupt appearance of the AX red shift amplitude ${R}^\text{A}_\text{plasma}$ at $n_\text{C}$. Moreover, it is substantiated by the jump of the lattice heating-induced red shift amplitudes $R^\text{A}_\text{heat}$ and $R^\text{B}_\text{heat}$ at $n_\text{C}$. The increased lattice heating can likely be attributed to an increased electron-phonon scattering cross section as free carriers have an increased mobility compared to excitons. 
It is important to note that the increase in the AX blue shift amplitude $B_\text{pop}^\text{A}$ over $n_\text{C}$ does not contradict the interpretation of a discontinuous Mott transition. As explained above, within the rate equation model the AX blue shift amplitude $B_\text{pop}^\text{A}$ associated with $\tau_\text{pop}$ is proportional to the total excitation density, which is composed of the sum of the plasma and exciton population and is independent of the dissociation ratio. It is therefore to be expected that it does not change abruptly at the critical density, and the fact that this is not the case implicitly shows that the assumption of the rate equation model must be satisfied, i.e., that the plasma decays through the combined action of exciton formation and recombination.

It can be noted that, compared to theoretical predictions of plasma-induced BGR at the Mott density, the values determined here are relatively low. The experimentally determined value of the BGR by this method is, taking the BX shift as reference, $-$70 meV which is considerably lower than the theoretically calculated value of about $-$350\ meV \cite{steinhoff2017exciton}. This is likely because the BGR is probed via an optical transition, instead of direct measurement of electronic energies via, e.g. photoelectron spectroscopy \cite{liu2019direct}. Screening leads to a concurrent blue shift of excitonic resonances due to the reduction of the binding energy as is known from previous studies of the excitonic regime  \cite{calati2023dynamic}. The two effects, binding energy reduction and BGR partially compensate \cite{champagne2023quasiparticle}, resulting in a smaller effective red shift. 

In summary of this section, both the AX and the BX exhibit an abrupt and strong red shift at the same excitation density of $n_\text{C}$~=~29~$\pm$~2~x~10$^{12}$~cm$^{-2}$, which decays with the same timescale. We attribute this to free carrier-induced BGR upon crossing the Mott density. A simple rate equation model is introduced, which assumes that the photoexcitation - besides heating the lattice - creates excitons, and for sufficiently high excitation densities, dissociated excitons, i.e., free carriers. The free carriers have two decay channels, namely recombination and exciton formation. This model describes the observed shift dynamics for both the AX and the BX very well for both the excitonic and plasma phase with globally determined time constants for the plasma decay, the population lifetime and the lattice cooling. The global fit is quantitatively consistent with independent fits performed in the excitonic regime. The fitting enables the tracking of the amplitudes of the free carrier-induced BGR, the lattice heating-induced BGR and the total pump-induced population on the excitonic resonances across all excitation densities. All these parameters (besides the total pump-induced population) show an abrupt change at the critical density.
\section{Discussion and Conclusion}
The above evaluation of the broadband differential TA data using a careful lineshape analysis has enabled us to discern that monolayer WS$_2$ exhibits two distinct regimes of photoexcited dynamics, which are separated by an abrupt change at a critical excitation density. At low excitation densities up to 26.5~x~10$^{12}$~cm$^{-2}$, the pump-induced dynamics are dominated by excitons and lattice heating, happening at their characteristic time scales. While lattice heating leads to a red shift through BGR that affects all excitonic resonances, the presence of A excitons uniquely impacts the AX peak position, causing it to blue shift. It does not affect the BX peak position, as theoretically predicted \cite{katsch2019theory}. This independence of the BX peak position from the A exciton population makes it a spectator of even small changes of the BGR, and the analysis shows that a small quantity of pump-induced free charge carriers must be present in the excitonic regime. At 31~x~10$^{12}$~cm$^{-2}$ and higher excitation densities, the optical spectra suddenly testify the presence of a large number of free charge carriers. The plasma creates BGR and red shifts the lowest optical transitions, the AX and the BX likewise. As the quasiparticle density is decreased in time, and rapidly falls below the Mott limit, the plasma decays gradually by A exciton formation and recombination, with a time constant of 0.65~ps.  
The re-emergence of the excitonic phase after the plasma decay is demonstrated by the recurrence of the dynamics of the excitonic phase, in particular the A exciton population-induced blue shift of the AX. 

The observation of two distinct regimes, separated by a critical excitation density, shows that the EMT in monolayer WS$_2$ is a sharp transition. Within an excitation density window of only 4.5 x 10$^{12}$ cm$^{-2}$ i.e., 15\% of the Mott density $n_\text{C}=(29~\pm$~2)~x~10$^{12}$~cm$^{-2}$, the shift amplitude of the B exciton increases by a factor of 10. The sharpness of the jump is comparable to the calculation by Steinhoff et al \cite{steinhoff2017exciton}. In both, the theoretical calculation of Ref.~\cite{steinhoff2017exciton} and our experiment, the sharpness is limited by the amount of data points across the critical density. This observation is in contrast to the gradual reduction of the AX oscillator strength over a broad excitation density regime observed by us and Ref.~\cite{chernikov2015population}. The gradual bleach is the result of a population-induced ground state bleach, combined with a finite exciton dissociation time as discussed below. The sharp transition strongly suggests that the EMT has a discontinuous character but it cannot be regarded as proof. We expect that even if the EMT is truly discontinuous, inhomogeneities within the excitation spot as well as experimental parameters such as the laser power stability would lead to a finite broadening of the jump. A finer sampling across the critical density would therefore likely not help to resolve the question fundamentally. It is noteworthy that, in our experiments, the pump spot is about the same size as the probe laser spot as mentioned in the experimental section. This creates a macroscopic inhomogeneity of the pump laser fluence within the probe spot and is expected to broaden an initially sharp transition -- the center of the Gaussian-shaped excitation spot would undergo an EMT at a lower pump laser fluence than the outer ring segments. Despite all these existing causes that would lead to an apparent broadening, our experiments show a rather sharp jump as outlined before. While we currently cannot estimate the broadening effects quantitatively, an intriguing possibility would be that the transition is in fact of avalanche-type. In an avalanche-type transition, one could expect that if a sufficient part of the photoexcited area (e.g. the center of the photoexcited spot) undergoes an EMT, it will assist in dissociating 
excitons present in the neighbouring areas that have not yet reached the Mott density \cite{Rabinovich2024}. To explore this further, spatial resolution could be a next step, e.g. monitoring the TA locally at the center of a Gaussian-shaped excitation spot, and comparing this to an outer segment.

Our observation of a sharp EMT is in agreement with the abrupt photoluminescence change under cw laser excitation at a critical density, observed in monolayer MoS$_2$ and attributed to the EMT \cite{bataller2019dense}. Apparently, W-based and Mo-based TMDC systems do not behave differently in this respect, despite theoretical predictions to do so due to their difference in exciton binding energy \cite{guerci2019exciton}. The finding of a sharp transition is in agreement with the material-realistic many-body theory \cite{steinhoff2017exciton}. The finite amount of plasma formed before the critical density resolved through the BX shift is explained by the exciton-screening induced dissociation by Steinhoff et al.~\cite{steinhoff2017exciton}. Plasma formation, locally within the excited oscillator strength, could however, also be favoured via exciton dissociation at defects \cite{handa2024spontaneous}. It can be noted that the Mott density $n_\text{C}$~=~29~$\pm$~2~x~10$^{12}$~cm$^{-2}$ is somewhat higher than the value of 8~x~10$^{12}$~cm$^{-2}$ predicted by Steinhoff et al. and is notably close to the critical density of 34~x~10$^{12}$~cm$^{-2}$ in the cw laser experiments in MoS$_2$. In their case they attributed this phase to an electron-hole liquid instead of an electron hole plasma \cite{yu2019room}. The existence of this exotic phase in this and other studies \cite{arp2019electron} is substantiated by imaging of the photoexcited spot. This once again motivates that the employment of spatial-resolved techniques together with ultrafast time resolution to the EMT in WS$_2$ could be an interesting topic for future studies.   

We observe this transition in combination with the previously observed gradual bleach of the A exciton resonance as a function of excitation density (cf. Fig.~\ref{Fig:1}(d)). As explained in section \ref{sec:data_and_excitation_density}, the A exciton bleach can be entirely explained by the ground state bleach due to the pump-induced population. It clearly shows no abrupt change at the Mott density, which means that there is no screening-induced change of the A exciton oscillator strength. The persistence of the A and B excitonic resonance above the Mott density shows that excitons are still eigenstates of the system. This seems controversial with our interpretation of a discontinuous Mott transition at first sight. However, as mentioned in the introduction, even if an incoherent exciton population is formed, this population would be expected to dissociate and form plasma only within a screening time scale. In our case, this must have happened within 200~fs, which is also in agreement with measurements for bulk WSe$_2$ \cite{dendzik2020observation}. These findings shows that simply measuring the oscillator strength of excitonic resonances as indicator of the EMT is not sufficient. However, their shift dynamics can be used to examine the nature of the transition and the associated time constants as we have shown. 

With this study, we demonstrate the applicability of optical spectroscopy combined with a detailed lineshape analysis of the differential spectra to study Mott transitions via excitonic resonances. The sharp character of the particular Mott transition in WS$_2$ monolayers could be useful for photoswitches, exploiting the change of the optical, or electronic properties. 
\begin{acknowledgments}
We acknowledge Xiaoyang Zhu's continuing support of the WS$_2$ project.
\end{acknowledgments}
\section*{Data availability}
The data analysed in the present study are publicly available under https://doi.org/10.5281/zenodo.23211020
\bibliography{references}

\begin{thebibliography}{58}%
\makeatletter
\providecommand \@ifxundefined [1]{%
 \@ifx{#1\undefined}
}%
\providecommand \@ifnum [1]{%
 \ifnum #1\expandafter \@firstoftwo
 \else \expandafter \@secondoftwo
 \fi
}%
\providecommand \@ifx [1]{%
 \ifx #1\expandafter \@firstoftwo
 \else \expandafter \@secondoftwo
 \fi
}%
\providecommand \natexlab [1]{#1}%
\providecommand \enquote  [1]{``#1''}%
\providecommand \bibnamefont  [1]{#1}%
\providecommand \bibfnamefont [1]{#1}%
\providecommand \citenamefont [1]{#1}%
\providecommand \href@noop [0]{\@secondoftwo}%
\providecommand \href [0]{\begingroup \@sanitize@url \@href}%
\providecommand \@href[1]{\@@startlink{#1}\@@href}%
\providecommand \@@href[1]{\endgroup#1\@@endlink}%
\providecommand \@sanitize@url [0]{\catcode `\\12\catcode `\$12\catcode
  `\&12\catcode `\#12\catcode `\^12\catcode `\_12\catcode `\%12\relax}%
\providecommand \@@startlink[1]{}%
\providecommand \@@endlink[0]{}%
\providecommand \url  [0]{\begingroup\@sanitize@url \@url }%
\providecommand \@url [1]{\endgroup\@href {#1}{\urlprefix }}%
\providecommand \urlprefix  [0]{URL }%
\providecommand \Eprint [0]{\href }%
\providecommand \doibase [0]{https://doi.org/}%
\providecommand \selectlanguage [0]{\@gobble}%
\providecommand \bibinfo  [0]{\@secondoftwo}%
\providecommand \bibfield  [0]{\@secondoftwo}%
\providecommand \translation [1]{[#1]}%
\providecommand \BibitemOpen [0]{}%
\providecommand \bibitemStop [0]{}%
\providecommand \bibitemNoStop [0]{.\EOS\space}%
\providecommand \EOS [0]{\spacefactor3000\relax}%
\providecommand \BibitemShut  [1]{\csname bibitem#1\endcsname}%
\let\auto@bib@innerbib\@empty
\bibitem [{\citenamefont {Ye}\ \emph {et~al.}(2017)\citenamefont {Ye},
  \citenamefont {Zhang},\ and\ \citenamefont {Yap}}]{ye2017recent}%
  \BibitemOpen
  \bibfield  {author} {\bibinfo {author} {\bibfnamefont {M.}~\bibnamefont
  {Ye}}, \bibinfo {author} {\bibfnamefont {D.}~\bibnamefont {Zhang}},\ and\
  \bibinfo {author} {\bibfnamefont {Y.~K.}\ \bibnamefont {Yap}},\ }\bibfield
  {title} {\bibinfo {title} {Recent advances in electronic and optoelectronic
  devices based on two-dimensional transition metal dichalcogenides},\ }\href
  {https://doi.org/10.3390/electronics6020043} {\bibfield  {journal} {\bibinfo
  {journal} {Electronics}\ }\textbf {\bibinfo {volume} {6}},\ \bibinfo {pages}
  {43} (\bibinfo {year} {2017})}\BibitemShut {NoStop}%
\bibitem [{\citenamefont {Liu}\ and\ \citenamefont {Hersam}(2019)}]{liu20192d}%
  \BibitemOpen
  \bibfield  {author} {\bibinfo {author} {\bibfnamefont {X.}~\bibnamefont
  {Liu}}\ and\ \bibinfo {author} {\bibfnamefont {M.~C.}\ \bibnamefont
  {Hersam}},\ }\bibfield  {title} {\bibinfo {title} {{2D} materials for quantum
  information science},\ }\href {https://doi.org/10.1038/s41578-019-0136-x}
  {\bibfield  {journal} {\bibinfo  {journal} {Nature Reviews Materials}\
  }\textbf {\bibinfo {volume} {4}},\ \bibinfo {pages} {669} (\bibinfo {year}
  {2019})}\BibitemShut {NoStop}%
\bibitem [{\citenamefont {Wu}\ \emph {et~al.}(2015)\citenamefont {Wu},
  \citenamefont {Buckley}, \citenamefont {Schaibley}, \citenamefont {Feng},
  \citenamefont {Yan}, \citenamefont {Mandrus}, \citenamefont {Hatami},
  \citenamefont {Yao}, \citenamefont {Vu{\v{c}}kovi{\'c}}, \citenamefont
  {Majumdar} \emph {et~al.}}]{wu2015monolayer}%
  \BibitemOpen
  \bibfield  {author} {\bibinfo {author} {\bibfnamefont {S.}~\bibnamefont
  {Wu}}, \bibinfo {author} {\bibfnamefont {S.}~\bibnamefont {Buckley}},
  \bibinfo {author} {\bibfnamefont {J.~R.}\ \bibnamefont {Schaibley}}, \bibinfo
  {author} {\bibfnamefont {L.}~\bibnamefont {Feng}}, \bibinfo {author}
  {\bibfnamefont {J.}~\bibnamefont {Yan}}, \bibinfo {author} {\bibfnamefont
  {D.~G.}\ \bibnamefont {Mandrus}}, \bibinfo {author} {\bibfnamefont
  {F.}~\bibnamefont {Hatami}}, \bibinfo {author} {\bibfnamefont
  {W.}~\bibnamefont {Yao}}, \bibinfo {author} {\bibfnamefont {J.}~\bibnamefont
  {Vu{\v{c}}kovi{\'c}}}, \bibinfo {author} {\bibfnamefont {A.}~\bibnamefont
  {Majumdar}}, \emph {et~al.},\ }\bibfield  {title} {\bibinfo {title}
  {Monolayer semiconductor nanocavity lasers with ultralow thresholds},\ }\href
  {https://doi.org/10.1038/nature14290} {\bibfield  {journal} {\bibinfo
  {journal} {Nature}\ }\textbf {\bibinfo {volume} {520}},\ \bibinfo {pages}
  {69} (\bibinfo {year} {2015})}\BibitemShut {NoStop}%
\bibitem [{\citenamefont {Wang}\ \emph {et~al.}(2019)\citenamefont {Wang},
  \citenamefont {Ardelean}, \citenamefont {Bai}, \citenamefont {Steinhoff},
  \citenamefont {Florian}, \citenamefont {Jahnke}, \citenamefont {Xu},
  \citenamefont {Kira}, \citenamefont {Hone},\ and\ \citenamefont
  {Zhu}}]{wang2019optical}%
  \BibitemOpen
  \bibfield  {author} {\bibinfo {author} {\bibfnamefont {J.}~\bibnamefont
  {Wang}}, \bibinfo {author} {\bibfnamefont {J.}~\bibnamefont {Ardelean}},
  \bibinfo {author} {\bibfnamefont {Y.}~\bibnamefont {Bai}}, \bibinfo {author}
  {\bibfnamefont {A.}~\bibnamefont {Steinhoff}}, \bibinfo {author}
  {\bibfnamefont {M.}~\bibnamefont {Florian}}, \bibinfo {author} {\bibfnamefont
  {F.}~\bibnamefont {Jahnke}}, \bibinfo {author} {\bibfnamefont
  {X.}~\bibnamefont {Xu}}, \bibinfo {author} {\bibfnamefont {M.}~\bibnamefont
  {Kira}}, \bibinfo {author} {\bibfnamefont {J.}~\bibnamefont {Hone}},\ and\
  \bibinfo {author} {\bibfnamefont {X.-Y.}\ \bibnamefont {Zhu}},\ }\bibfield
  {title} {\bibinfo {title} {Optical generation of high carrier densities in
  {2D} semiconductor heterobilayers},\ }\href
  {https://doi.org/10.1126/sciadv.aax0145} {\bibfield  {journal} {\bibinfo
  {journal} {Science Advances}\ }\textbf {\bibinfo {volume} {5}},\ \bibinfo
  {pages} {eaax0145} (\bibinfo {year} {2019})}\BibitemShut {NoStop}%
\bibitem [{\citenamefont {Li}\ \emph {et~al.}(2021)\citenamefont {Li},
  \citenamefont {Sui}, \citenamefont {Niu}, \citenamefont {Jiang},
  \citenamefont {Zhang}, \citenamefont {Che}, \citenamefont {Wu}, \citenamefont
  {Jin},\ and\ \citenamefont {Yuan}}]{li2021ultrafast}%
  \BibitemOpen
  \bibfield  {author} {\bibinfo {author} {\bibfnamefont {Q.}~\bibnamefont
  {Li}}, \bibinfo {author} {\bibfnamefont {L.}~\bibnamefont {Sui}}, \bibinfo
  {author} {\bibfnamefont {G.}~\bibnamefont {Niu}}, \bibinfo {author}
  {\bibfnamefont {J.}~\bibnamefont {Jiang}}, \bibinfo {author} {\bibfnamefont
  {Y.}~\bibnamefont {Zhang}}, \bibinfo {author} {\bibfnamefont
  {L.}~\bibnamefont {Che}}, \bibinfo {author} {\bibfnamefont {G.}~\bibnamefont
  {Wu}}, \bibinfo {author} {\bibfnamefont {M.}~\bibnamefont {Jin}},\ and\
  \bibinfo {author} {\bibfnamefont {K.}~\bibnamefont {Yuan}},\ }\bibfield
  {title} {\bibinfo {title} {Ultrafast optical switching to a metallic state
  via photoinduced {Mott} transition in few-layer {MoS$_2$} under hydrostatic
  pressure},\ }\href {https://doi.org/10.1103/PhysRevB.103.125416} {\bibfield
  {journal} {\bibinfo  {journal} {Physical Review B}\ }\textbf {\bibinfo
  {volume} {103}},\ \bibinfo {pages} {125416} (\bibinfo {year}
  {2021})}\BibitemShut {NoStop}%
\bibitem [{\citenamefont {Yu}\ \emph {et~al.}(2020)\citenamefont {Yu},
  \citenamefont {Li},\ and\ \citenamefont {Cao}}]{yu2020exciton}%
  \BibitemOpen
  \bibfield  {author} {\bibinfo {author} {\bibfnamefont {Y.}~\bibnamefont
  {Yu}}, \bibinfo {author} {\bibfnamefont {G.}~\bibnamefont {Li}},\ and\
  \bibinfo {author} {\bibfnamefont {L.}~\bibnamefont {Cao}},\ }\bibfield
  {title} {\bibinfo {title} {Exciton {Mott} transition in two-dimensional
  semiconductors},\ }\bibfield  {journal} {\bibinfo  {journal} {arXiv preprint
  arXiv:2007.11509}\ }\href {https://doi.org/10.48550/arXiv.2007.11509}
  {10.48550/arXiv.2007.11509} (\bibinfo {year} {2020})\BibitemShut {NoStop}%
\bibitem [{\citenamefont {Xu}\ \emph {et~al.}(2025)\citenamefont {Xu},
  \citenamefont {Xiang}, \citenamefont {Shi}, \citenamefont {Zhai},
  \citenamefont {Dai}, \citenamefont {Wang}, \citenamefont {Liu}, \citenamefont
  {Yu},\ and\ \citenamefont {He}}]{xu2025room}%
  \BibitemOpen
  \bibfield  {author} {\bibinfo {author} {\bibfnamefont {Y.}~\bibnamefont
  {Xu}}, \bibinfo {author} {\bibfnamefont {Y.}~\bibnamefont {Xiang}}, \bibinfo
  {author} {\bibfnamefont {M.}~\bibnamefont {Shi}}, \bibinfo {author}
  {\bibfnamefont {B.}~\bibnamefont {Zhai}}, \bibinfo {author} {\bibfnamefont
  {W.}~\bibnamefont {Dai}}, \bibinfo {author} {\bibfnamefont {T.}~\bibnamefont
  {Wang}}, \bibinfo {author} {\bibfnamefont {X.}~\bibnamefont {Liu}}, \bibinfo
  {author} {\bibfnamefont {Y.}~\bibnamefont {Yu}},\ and\ \bibinfo {author}
  {\bibfnamefont {J.}~\bibnamefont {He}},\ }\bibfield  {title} {\bibinfo
  {title} {Room-temperature amplified spontaneous emission in two-dimensional
  {WS$_2$} beyond exciton {Mott} transition},\ }\href
  {https://doi.org/10.1103/PhysRevLett.134.066904} {\bibfield  {journal}
  {\bibinfo  {journal} {Physical Review Letters}\ }\textbf {\bibinfo {volume}
  {134}},\ \bibinfo {pages} {066904} (\bibinfo {year} {2025})}\BibitemShut
  {NoStop}%
\bibitem [{\citenamefont {Siday}\ \emph {et~al.}(2022)\citenamefont {Siday},
  \citenamefont {Sandner}, \citenamefont {Brem}, \citenamefont {Zizlsperger},
  \citenamefont {Perea-Causin}, \citenamefont {Schiegl}, \citenamefont
  {Nerreter}, \citenamefont {Plankl}, \citenamefont {Merkl}, \citenamefont
  {Mooshammer} \emph {et~al.}}]{siday2022ultrafast}%
  \BibitemOpen
  \bibfield  {author} {\bibinfo {author} {\bibfnamefont {T.}~\bibnamefont
  {Siday}}, \bibinfo {author} {\bibfnamefont {F.}~\bibnamefont {Sandner}},
  \bibinfo {author} {\bibfnamefont {S.}~\bibnamefont {Brem}}, \bibinfo {author}
  {\bibfnamefont {M.}~\bibnamefont {Zizlsperger}}, \bibinfo {author}
  {\bibfnamefont {R.}~\bibnamefont {Perea-Causin}}, \bibinfo {author}
  {\bibfnamefont {F.}~\bibnamefont {Schiegl}}, \bibinfo {author} {\bibfnamefont
  {S.}~\bibnamefont {Nerreter}}, \bibinfo {author} {\bibfnamefont
  {M.}~\bibnamefont {Plankl}}, \bibinfo {author} {\bibfnamefont
  {P.}~\bibnamefont {Merkl}}, \bibinfo {author} {\bibfnamefont
  {F.}~\bibnamefont {Mooshammer}}, \emph {et~al.},\ }\bibfield  {title}
  {\bibinfo {title} {Ultrafast nanoscopy of high-density exciton phases in
  {WSe$_2$}},\ }\href {https://doi.org/10.1021/acs.nanolett.1c04741} {\bibfield
   {journal} {\bibinfo  {journal} {Nano Letters}\ }\textbf {\bibinfo {volume}
  {22}},\ \bibinfo {pages} {2561} (\bibinfo {year} {2022})}\BibitemShut
  {NoStop}%
\bibitem [{\citenamefont {Pekh}\ \emph {et~al.}(2020)\citenamefont {Pekh},
  \citenamefont {Ratnikov},\ and\ \citenamefont {Silin}}]{pekh2020electron}%
  \BibitemOpen
  \bibfield  {author} {\bibinfo {author} {\bibfnamefont {P.~L.}\ \bibnamefont
  {Pekh}}, \bibinfo {author} {\bibfnamefont {P.~V.}\ \bibnamefont {Ratnikov}},\
  and\ \bibinfo {author} {\bibfnamefont {A.~P.}\ \bibnamefont {Silin}},\
  }\bibfield  {title} {\bibinfo {title} {Electron-hole liquid in monolayer
  transition metal dichalcogenide heterostructures},\ }\href
  {https://doi.org//10.1134/S0021364020020101} {\bibfield  {journal} {\bibinfo
  {journal} {JETP Letters}\ }\textbf {\bibinfo {volume} {111}},\ \bibinfo
  {pages} {90} (\bibinfo {year} {2020})}\BibitemShut {NoStop}%
\bibitem [{\citenamefont {Bataller}\ \emph {et~al.}(2019)\citenamefont
  {Bataller}, \citenamefont {Younts}, \citenamefont {Rustagi}, \citenamefont
  {Yu}, \citenamefont {Ardekani}, \citenamefont {Kemper}, \citenamefont {Cao},\
  and\ \citenamefont {Gundogdu}}]{bataller2019dense}%
  \BibitemOpen
  \bibfield  {author} {\bibinfo {author} {\bibfnamefont {A.~W.}\ \bibnamefont
  {Bataller}}, \bibinfo {author} {\bibfnamefont {R.~A.}\ \bibnamefont
  {Younts}}, \bibinfo {author} {\bibfnamefont {A.}~\bibnamefont {Rustagi}},
  \bibinfo {author} {\bibfnamefont {Y.}~\bibnamefont {Yu}}, \bibinfo {author}
  {\bibfnamefont {H.}~\bibnamefont {Ardekani}}, \bibinfo {author}
  {\bibfnamefont {A.}~\bibnamefont {Kemper}}, \bibinfo {author} {\bibfnamefont
  {L.}~\bibnamefont {Cao}},\ and\ \bibinfo {author} {\bibfnamefont
  {K.}~\bibnamefont {Gundogdu}},\ }\bibfield  {title} {\bibinfo {title} {Dense
  electron--hole plasma formation and ultralong charge lifetime in monolayer
  {MoS$_2$} via material tuning},\ }\href
  {https://doi.org/10.1021/acs.nanolett.8b04408} {\bibfield  {journal}
  {\bibinfo  {journal} {Nano Letters}\ }\textbf {\bibinfo {volume} {19}},\
  \bibinfo {pages} {1104} (\bibinfo {year} {2019})}\BibitemShut {NoStop}%
\bibitem [{\citenamefont {Sousa}\ \emph {et~al.}(2023)\citenamefont {Sousa},
  \citenamefont {Perea-Causin}, \citenamefont {Hartmann}, \citenamefont
  {Lafet{\'a}}, \citenamefont {Rosa}, \citenamefont {Brem}, \citenamefont
  {Palekar}, \citenamefont {Reitzenstein}, \citenamefont {Hartschuh},
  \citenamefont {Malic} \emph {et~al.}}]{sousa2023ultrafast}%
  \BibitemOpen
  \bibfield  {author} {\bibinfo {author} {\bibfnamefont {F.~B.}\ \bibnamefont
  {Sousa}}, \bibinfo {author} {\bibfnamefont {R.}~\bibnamefont {Perea-Causin}},
  \bibinfo {author} {\bibfnamefont {S.}~\bibnamefont {Hartmann}}, \bibinfo
  {author} {\bibfnamefont {L.}~\bibnamefont {Lafet{\'a}}}, \bibinfo {author}
  {\bibfnamefont {B.}~\bibnamefont {Rosa}}, \bibinfo {author} {\bibfnamefont
  {S.}~\bibnamefont {Brem}}, \bibinfo {author} {\bibfnamefont {C.}~\bibnamefont
  {Palekar}}, \bibinfo {author} {\bibfnamefont {S.}~\bibnamefont
  {Reitzenstein}}, \bibinfo {author} {\bibfnamefont {A.}~\bibnamefont
  {Hartschuh}}, \bibinfo {author} {\bibfnamefont {E.}~\bibnamefont {Malic}},
  \emph {et~al.},\ }\bibfield  {title} {\bibinfo {title} {Ultrafast hot
  electron--hole plasma photoluminescence in two-dimensional semiconductors},\
  }\href {https://doi.org/10.1039/d2nr06732c} {\bibfield  {journal} {\bibinfo
  {journal} {Nanoscale}\ }\textbf {\bibinfo {volume} {15}},\ \bibinfo {pages}
  {7154} (\bibinfo {year} {2023})}\BibitemShut {NoStop}%
\bibitem [{\citenamefont {Chernikov}\ \emph {et~al.}(2015)\citenamefont
  {Chernikov}, \citenamefont {Ruppert}, \citenamefont {Hill}, \citenamefont
  {Rigosi},\ and\ \citenamefont {Heinz}}]{chernikov2015population}%
  \BibitemOpen
  \bibfield  {author} {\bibinfo {author} {\bibfnamefont {A.}~\bibnamefont
  {Chernikov}}, \bibinfo {author} {\bibfnamefont {C.}~\bibnamefont {Ruppert}},
  \bibinfo {author} {\bibfnamefont {H.~M.}\ \bibnamefont {Hill}}, \bibinfo
  {author} {\bibfnamefont {A.~F.}\ \bibnamefont {Rigosi}},\ and\ \bibinfo
  {author} {\bibfnamefont {T.~F.}\ \bibnamefont {Heinz}},\ }\bibfield  {title}
  {\bibinfo {title} {Population inversion and giant bandgap renormalization in
  atomically thin {WS$_2$} layers},\ }\href
  {https://doi.org/10.1038/nphoton.2015.104} {\bibfield  {journal} {\bibinfo
  {journal} {Nature Photonics}\ }\textbf {\bibinfo {volume} {9}},\ \bibinfo
  {pages} {466} (\bibinfo {year} {2015})}\BibitemShut {NoStop}%
\bibitem [{\citenamefont {Steinhoff}\ \emph {et~al.}(2017)\citenamefont
  {Steinhoff}, \citenamefont {Florian}, \citenamefont {R{\"o}sner},
  \citenamefont {Sch{\"o}nhoff}, \citenamefont {Wehling},\ and\ \citenamefont
  {Jahnke}}]{steinhoff2017exciton}%
  \BibitemOpen
  \bibfield  {author} {\bibinfo {author} {\bibfnamefont {A.}~\bibnamefont
  {Steinhoff}}, \bibinfo {author} {\bibfnamefont {M.}~\bibnamefont {Florian}},
  \bibinfo {author} {\bibfnamefont {M.}~\bibnamefont {R{\"o}sner}}, \bibinfo
  {author} {\bibfnamefont {G.}~\bibnamefont {Sch{\"o}nhoff}}, \bibinfo {author}
  {\bibfnamefont {T.~O.}\ \bibnamefont {Wehling}},\ and\ \bibinfo {author}
  {\bibfnamefont {F.}~\bibnamefont {Jahnke}},\ }\bibfield  {title} {\bibinfo
  {title} {Exciton fission in monolayer transition metal dichalcogenide
  semiconductors},\ }\href {https://doi.org/10.1038/s41467-017-01298-6}
  {\bibfield  {journal} {\bibinfo  {journal} {Nature Communications}\ }\textbf
  {\bibinfo {volume} {8}},\ \bibinfo {pages} {1166} (\bibinfo {year}
  {2017})}\BibitemShut {NoStop}%
\bibitem [{\citenamefont {Guerci}\ \emph {et~al.}(2019)\citenamefont {Guerci},
  \citenamefont {Capone},\ and\ \citenamefont {Fabrizio}}]{guerci2019exciton}%
  \BibitemOpen
  \bibfield  {author} {\bibinfo {author} {\bibfnamefont {D.}~\bibnamefont
  {Guerci}}, \bibinfo {author} {\bibfnamefont {M.}~\bibnamefont {Capone}},\
  and\ \bibinfo {author} {\bibfnamefont {M.}~\bibnamefont {Fabrizio}},\
  }\bibfield  {title} {\bibinfo {title} {Exciton {Mott} transition revisited},\
  }\href {https://doi.org/10.1103/PhysRevMaterials.3.054605} {\bibfield
  {journal} {\bibinfo  {journal} {Physical Review Materials}\ }\textbf
  {\bibinfo {volume} {3}},\ \bibinfo {pages} {054605} (\bibinfo {year}
  {2019})}\BibitemShut {NoStop}%
\bibitem [{\citenamefont {Handa}\ \emph {et~al.}(2024)\citenamefont {Handa},
  \citenamefont {Holbrook}, \citenamefont {Olsen}, \citenamefont {Holtzman},
  \citenamefont {Huber}, \citenamefont {Wang}, \citenamefont {Bonn},
  \citenamefont {Barmak}, \citenamefont {Hone}, \citenamefont {Pasupathy} \emph
  {et~al.}}]{handa2024spontaneous}%
  \BibitemOpen
  \bibfield  {author} {\bibinfo {author} {\bibfnamefont {T.}~\bibnamefont
  {Handa}}, \bibinfo {author} {\bibfnamefont {M.}~\bibnamefont {Holbrook}},
  \bibinfo {author} {\bibfnamefont {N.}~\bibnamefont {Olsen}}, \bibinfo
  {author} {\bibfnamefont {L.~N.}\ \bibnamefont {Holtzman}}, \bibinfo {author}
  {\bibfnamefont {L.}~\bibnamefont {Huber}}, \bibinfo {author} {\bibfnamefont
  {H.~I.}\ \bibnamefont {Wang}}, \bibinfo {author} {\bibfnamefont
  {M.}~\bibnamefont {Bonn}}, \bibinfo {author} {\bibfnamefont {K.}~\bibnamefont
  {Barmak}}, \bibinfo {author} {\bibfnamefont {J.~C.}\ \bibnamefont {Hone}},
  \bibinfo {author} {\bibfnamefont {A.~N.}\ \bibnamefont {Pasupathy}}, \emph
  {et~al.},\ }\bibfield  {title} {\bibinfo {title} {Spontaneous exciton
  dissociation in transition metal dichalcogenide monolayers},\ }\href
  {https://doi.org/10.1126/sciadv.adj4060} {\bibfield  {journal} {\bibinfo
  {journal} {Science Advances}\ }\textbf {\bibinfo {volume} {10}},\ \bibinfo
  {pages} {eadj4060} (\bibinfo {year} {2024})}\BibitemShut {NoStop}%
\bibitem [{\citenamefont {Yu}\ \emph {et~al.}(2019)\citenamefont {Yu},
  \citenamefont {Bataller}, \citenamefont {Younts}, \citenamefont {Yu},
  \citenamefont {Li}, \citenamefont {Puretzky}, \citenamefont {Geohegan},
  \citenamefont {Gundogdu},\ and\ \citenamefont {Cao}}]{yu2019room}%
  \BibitemOpen
  \bibfield  {author} {\bibinfo {author} {\bibfnamefont {Y.}~\bibnamefont
  {Yu}}, \bibinfo {author} {\bibfnamefont {A.~W.}\ \bibnamefont {Bataller}},
  \bibinfo {author} {\bibfnamefont {R.}~\bibnamefont {Younts}}, \bibinfo
  {author} {\bibfnamefont {Y.}~\bibnamefont {Yu}}, \bibinfo {author}
  {\bibfnamefont {G.}~\bibnamefont {Li}}, \bibinfo {author} {\bibfnamefont
  {A.~A.}\ \bibnamefont {Puretzky}}, \bibinfo {author} {\bibfnamefont {D.~B.}\
  \bibnamefont {Geohegan}}, \bibinfo {author} {\bibfnamefont {K.}~\bibnamefont
  {Gundogdu}},\ and\ \bibinfo {author} {\bibfnamefont {L.}~\bibnamefont
  {Cao}},\ }\bibfield  {title} {\bibinfo {title} {Room-temperature
  electron--hole liquid in monolayer {MoS$_2$}},\ }\href
  {https://doi.org/10.1021/acsnano.9b04124} {\bibfield  {journal} {\bibinfo
  {journal} {ACS Nano}\ }\textbf {\bibinfo {volume} {13}},\ \bibinfo {pages}
  {10351} (\bibinfo {year} {2019})}\BibitemShut {NoStop}%
\bibitem [{\citenamefont {Arp}\ \emph {et~al.}(2019)\citenamefont {Arp},
  \citenamefont {Pleskot}, \citenamefont {Aji},\ and\ \citenamefont
  {Gabor}}]{arp2019electron}%
  \BibitemOpen
  \bibfield  {author} {\bibinfo {author} {\bibfnamefont {T.~B.}\ \bibnamefont
  {Arp}}, \bibinfo {author} {\bibfnamefont {D.}~\bibnamefont {Pleskot}},
  \bibinfo {author} {\bibfnamefont {V.}~\bibnamefont {Aji}},\ and\ \bibinfo
  {author} {\bibfnamefont {N.~M.}\ \bibnamefont {Gabor}},\ }\bibfield  {title}
  {\bibinfo {title} {Electron--hole liquid in a van der {W}aals heterostructure
  photocell at room temperature},\ }\href
  {https://doi.org/10.1038/s41566-019-0349-y} {\bibfield  {journal} {\bibinfo
  {journal} {Nature Photonics}\ }\textbf {\bibinfo {volume} {13}},\ \bibinfo
  {pages} {245} (\bibinfo {year} {2019})}\BibitemShut {NoStop}%
\bibitem [{\citenamefont {Rabinovich}\ \emph {et~al.}(2025)\citenamefont
  {Rabinovich}, \citenamefont {Yaresko}, \citenamefont {Dawson}, \citenamefont
  {Krautloher}, \citenamefont {Priessnitz}, \citenamefont {Mathis},
  \citenamefont {Kirilyuk}, \citenamefont {Keimer},\ and\ \citenamefont
  {Boris}}]{Rabinovich2024}%
  \BibitemOpen
  \bibfield  {author} {\bibinfo {author} {\bibfnamefont {K.~S.}\ \bibnamefont
  {Rabinovich}}, \bibinfo {author} {\bibfnamefont {A.~N.}\ \bibnamefont
  {Yaresko}}, \bibinfo {author} {\bibfnamefont {R.~D.}\ \bibnamefont {Dawson}},
  \bibinfo {author} {\bibfnamefont {M.~J.}\ \bibnamefont {Krautloher}},
  \bibinfo {author} {\bibfnamefont {T.}~\bibnamefont {Priessnitz}}, \bibinfo
  {author} {\bibfnamefont {Y.-L.}\ \bibnamefont {Mathis}}, \bibinfo {author}
  {\bibfnamefont {A.}~\bibnamefont {Kirilyuk}}, \bibinfo {author}
  {\bibfnamefont {B.}~\bibnamefont {Keimer}},\ and\ \bibinfo {author}
  {\bibfnamefont {A.~V.}\ \bibnamefont {Boris}},\ }\bibfield  {title} {\bibinfo
  {title} {Optically driven dynamics of a mott insulator-to-metal transition},\
  }\href {https://doi.org/10.1002/adfm.202416597} {\bibfield  {journal}
  {\bibinfo  {journal} {Advanced Functional Materials}\ }\textbf {\bibinfo
  {volume} {35}},\ \bibinfo {pages} {2416597} (\bibinfo {year}
  {2025})}\BibitemShut {NoStop}%
\bibitem [{\citenamefont {Dendzik}\ \emph {et~al.}(2020)\citenamefont
  {Dendzik}, \citenamefont {Xian}, \citenamefont {Perfetto}, \citenamefont
  {Sangalli}, \citenamefont {Kutnyakhov}, \citenamefont {Dong}, \citenamefont
  {Beaulieu}, \citenamefont {Pincelli}, \citenamefont {Pressacco},
  \citenamefont {Curcio} \emph {et~al.}}]{dendzik2020observation}%
  \BibitemOpen
  \bibfield  {author} {\bibinfo {author} {\bibfnamefont {M.}~\bibnamefont
  {Dendzik}}, \bibinfo {author} {\bibfnamefont {R.~P.}\ \bibnamefont {Xian}},
  \bibinfo {author} {\bibfnamefont {E.}~\bibnamefont {Perfetto}}, \bibinfo
  {author} {\bibfnamefont {D.}~\bibnamefont {Sangalli}}, \bibinfo {author}
  {\bibfnamefont {D.}~\bibnamefont {Kutnyakhov}}, \bibinfo {author}
  {\bibfnamefont {S.}~\bibnamefont {Dong}}, \bibinfo {author} {\bibfnamefont
  {S.}~\bibnamefont {Beaulieu}}, \bibinfo {author} {\bibfnamefont
  {T.}~\bibnamefont {Pincelli}}, \bibinfo {author} {\bibfnamefont
  {F.}~\bibnamefont {Pressacco}}, \bibinfo {author} {\bibfnamefont
  {D.}~\bibnamefont {Curcio}}, \emph {et~al.},\ }\bibfield  {title} {\bibinfo
  {title} {Observation of an excitonic {Mott} transition through ultrafast
  core-cum-conduction photoemission spectroscopy},\ }\href
  {https://doi.org/10.1103/PhysRevLett.125.096401} {\bibfield  {journal}
  {\bibinfo  {journal} {Physical Review Letters}\ }\textbf {\bibinfo {volume}
  {125}},\ \bibinfo {pages} {096401} (\bibinfo {year} {2020})}\BibitemShut
  {NoStop}%
\bibitem [{\citenamefont {Karmakar}\ \emph {et~al.}(2021)\citenamefont
  {Karmakar}, \citenamefont {Mukherjee}, \citenamefont {Ray},\ and\
  \citenamefont {Datta}}]{karmakar2021electron}%
  \BibitemOpen
  \bibfield  {author} {\bibinfo {author} {\bibfnamefont {M.}~\bibnamefont
  {Karmakar}}, \bibinfo {author} {\bibfnamefont {S.}~\bibnamefont {Mukherjee}},
  \bibinfo {author} {\bibfnamefont {S.~K.}\ \bibnamefont {Ray}},\ and\ \bibinfo
  {author} {\bibfnamefont {P.~K.}\ \bibnamefont {Datta}},\ }\bibfield  {title}
  {\bibinfo {title} {Electron-hole plasma formation dynamics observed through
  exciton-plasma interactions in transition metal dichalcogenides},\ }\href
  {https://doi.org/10.1103/PhysRevB.104.075446} {\bibfield  {journal} {\bibinfo
   {journal} {Physical Review B}\ }\textbf {\bibinfo {volume} {104}},\ \bibinfo
  {pages} {075446} (\bibinfo {year} {2021})}\BibitemShut {NoStop}%
\bibitem [{\citenamefont {Huber}\ \emph {et~al.}(2001)\citenamefont {Huber},
  \citenamefont {Tauser}, \citenamefont {Brodschelm}, \citenamefont {Bichler},
  \citenamefont {Abstreiter},\ and\ \citenamefont
  {Leitenstorfer}}]{huber2001many}%
  \BibitemOpen
  \bibfield  {author} {\bibinfo {author} {\bibfnamefont {R.}~\bibnamefont
  {Huber}}, \bibinfo {author} {\bibfnamefont {F.}~\bibnamefont {Tauser}},
  \bibinfo {author} {\bibfnamefont {A.}~\bibnamefont {Brodschelm}}, \bibinfo
  {author} {\bibfnamefont {M.}~\bibnamefont {Bichler}}, \bibinfo {author}
  {\bibfnamefont {G.}~\bibnamefont {Abstreiter}},\ and\ \bibinfo {author}
  {\bibfnamefont {A.}~\bibnamefont {Leitenstorfer}},\ }\bibfield  {title}
  {\bibinfo {title} {How many-particle interactions develop after ultrafast
  excitation of an electron--hole plasma},\ }\href
  {https://doi.org/10.1038/35104522} {\bibfield  {journal} {\bibinfo  {journal}
  {Nature}\ }\textbf {\bibinfo {volume} {414}},\ \bibinfo {pages} {286}
  (\bibinfo {year} {2001})}\BibitemShut {NoStop}%
\bibitem [{\citenamefont {Sch{\"o}ne}\ and\ \citenamefont
  {Ekardt}(2000)}]{schone2000time}%
  \BibitemOpen
  \bibfield  {author} {\bibinfo {author} {\bibfnamefont {W.-D.}\ \bibnamefont
  {Sch{\"o}ne}}\ and\ \bibinfo {author} {\bibfnamefont {W.}~\bibnamefont
  {Ekardt}},\ }\bibfield  {title} {\bibinfo {title} {Time-dependent screening
  of a positive charge distribution in metals: Excitons on an ultrashort time
  scale},\ }\href {https://doi.org/10.1103/PhysRevB.62.13464} {\bibfield
  {journal} {\bibinfo  {journal} {Physical Review B}\ }\textbf {\bibinfo
  {volume} {62}},\ \bibinfo {pages} {13464} (\bibinfo {year}
  {2000})}\BibitemShut {NoStop}%
\bibitem [{\citenamefont {Cui}\ \emph {et~al.}(2014)\citenamefont {Cui},
  \citenamefont {Ceballos}, \citenamefont {Kumar},\ and\ \citenamefont
  {Zhao}}]{cui2014transient}%
  \BibitemOpen
  \bibfield  {author} {\bibinfo {author} {\bibfnamefont {Q.}~\bibnamefont
  {Cui}}, \bibinfo {author} {\bibfnamefont {F.}~\bibnamefont {Ceballos}},
  \bibinfo {author} {\bibfnamefont {N.}~\bibnamefont {Kumar}},\ and\ \bibinfo
  {author} {\bibfnamefont {H.}~\bibnamefont {Zhao}},\ }\bibfield  {title}
  {\bibinfo {title} {Transient absorption microscopy of monolayer and bulk
  {WSe$_2$}},\ }\href {https://doi.org/10.1021/nn500277y} {\bibfield  {journal}
  {\bibinfo  {journal} {ACS Nano}\ }\textbf {\bibinfo {volume} {8}},\ \bibinfo
  {pages} {2970} (\bibinfo {year} {2014})}\BibitemShut {NoStop}%
\bibitem [{\citenamefont {Calati}\ \emph {et~al.}(2023)\citenamefont {Calati},
  \citenamefont {Li}, \citenamefont {Zhu},\ and\ \citenamefont
  {St{\"a}hler}}]{calati2023dynamic}%
  \BibitemOpen
  \bibfield  {author} {\bibinfo {author} {\bibfnamefont {S.}~\bibnamefont
  {Calati}}, \bibinfo {author} {\bibfnamefont {Q.}~\bibnamefont {Li}}, \bibinfo
  {author} {\bibfnamefont {X.}~\bibnamefont {Zhu}},\ and\ \bibinfo {author}
  {\bibfnamefont {J.}~\bibnamefont {St{\"a}hler}},\ }\bibfield  {title}
  {\bibinfo {title} {Dynamic screening of quasiparticles in {WS$_2$}
  monolayers},\ }\href {https://doi.org/10.1103/PhysRevB.107.115404} {\bibfield
   {journal} {\bibinfo  {journal} {Physical Review B}\ }\textbf {\bibinfo
  {volume} {107}},\ \bibinfo {pages} {115404} (\bibinfo {year}
  {2023})}\BibitemShut {NoStop}%
\bibitem [{\citenamefont {Ruppert}\ \emph {et~al.}(2017)\citenamefont
  {Ruppert}, \citenamefont {Chernikov}, \citenamefont {Hill}, \citenamefont
  {Rigosi},\ and\ \citenamefont {Heinz}}]{ruppert2017role}%
  \BibitemOpen
  \bibfield  {author} {\bibinfo {author} {\bibfnamefont {C.}~\bibnamefont
  {Ruppert}}, \bibinfo {author} {\bibfnamefont {A.}~\bibnamefont {Chernikov}},
  \bibinfo {author} {\bibfnamefont {H.~M.}\ \bibnamefont {Hill}}, \bibinfo
  {author} {\bibfnamefont {A.~F.}\ \bibnamefont {Rigosi}},\ and\ \bibinfo
  {author} {\bibfnamefont {T.~F.}\ \bibnamefont {Heinz}},\ }\bibfield  {title}
  {\bibinfo {title} {The role of electronic and phononic excitation in the
  optical response of monolayer {WS$_2$} after ultrafast excitation},\ }\href
  {https://doi.org/10.1021/acs.nanolett.6b03513} {\bibfield  {journal}
  {\bibinfo  {journal} {Nano Letters}\ }\textbf {\bibinfo {volume} {17}},\
  \bibinfo {pages} {644} (\bibinfo {year} {2017})}\BibitemShut {NoStop}%
\bibitem [{\citenamefont {Cunningham}\ \emph {et~al.}(2017)\citenamefont
  {Cunningham}, \citenamefont {Hanbicki}, \citenamefont {McCreary},\ and\
  \citenamefont {Jonker}}]{cunningham2017photoinduced}%
  \BibitemOpen
  \bibfield  {author} {\bibinfo {author} {\bibfnamefont {P.~D.}\ \bibnamefont
  {Cunningham}}, \bibinfo {author} {\bibfnamefont {A.~T.}\ \bibnamefont
  {Hanbicki}}, \bibinfo {author} {\bibfnamefont {K.~M.}\ \bibnamefont
  {McCreary}},\ and\ \bibinfo {author} {\bibfnamefont {B.~T.}\ \bibnamefont
  {Jonker}},\ }\bibfield  {title} {\bibinfo {title} {Photoinduced bandgap
  renormalization and exciton binding energy reduction in {WS$_2$}},\ }\href
  {https://doi.org/10.1021/acsnano.7b06885} {\bibfield  {journal} {\bibinfo
  {journal} {ACS Nano}\ }\textbf {\bibinfo {volume} {11}},\ \bibinfo {pages}
  {12601} (\bibinfo {year} {2017})}\BibitemShut {NoStop}%
\bibitem [{\citenamefont {Sie}\ \emph {et~al.}(2017)\citenamefont {Sie},
  \citenamefont {Steinhoff}, \citenamefont {Gies}, \citenamefont {Lui},
  \citenamefont {Ma}, \citenamefont {R{\"o}sner}, \citenamefont
  {Sch{\"o}nhoff}, \citenamefont {Jahnke}, \citenamefont {Wehling},
  \citenamefont {Lee} \emph {et~al.}}]{sie2017observation}%
  \BibitemOpen
  \bibfield  {author} {\bibinfo {author} {\bibfnamefont {E.~J.}\ \bibnamefont
  {Sie}}, \bibinfo {author} {\bibfnamefont {A.}~\bibnamefont {Steinhoff}},
  \bibinfo {author} {\bibfnamefont {C.}~\bibnamefont {Gies}}, \bibinfo {author}
  {\bibfnamefont {C.~H.}\ \bibnamefont {Lui}}, \bibinfo {author} {\bibfnamefont
  {Q.}~\bibnamefont {Ma}}, \bibinfo {author} {\bibfnamefont {M.}~\bibnamefont
  {R{\"o}sner}}, \bibinfo {author} {\bibfnamefont {G.}~\bibnamefont
  {Sch{\"o}nhoff}}, \bibinfo {author} {\bibfnamefont {F.}~\bibnamefont
  {Jahnke}}, \bibinfo {author} {\bibfnamefont {T.~O.}\ \bibnamefont {Wehling}},
  \bibinfo {author} {\bibfnamefont {Y.-H.}\ \bibnamefont {Lee}}, \emph
  {et~al.},\ }\bibfield  {title} {\bibinfo {title} {Observation of exciton
  redshift--blueshift crossover in monolayer {WS$_2$}},\ }\href
  {https://doi.org/10.1021/acs.nanolett.7b01034} {\bibfield  {journal}
  {\bibinfo  {journal} {Nano Letters}\ }\textbf {\bibinfo {volume} {17}},\
  \bibinfo {pages} {4210} (\bibinfo {year} {2017})}\BibitemShut {NoStop}%
\bibitem [{\citenamefont {Zou}\ \emph {et~al.}(2023)\citenamefont {Zou},
  \citenamefont {Zhu}, \citenamefont {Wang}, \citenamefont {Meng},
  \citenamefont {Wang}, \citenamefont {Chen},\ and\ \citenamefont
  {Weng}}]{zou2023coherent}%
  \BibitemOpen
  \bibfield  {author} {\bibinfo {author} {\bibfnamefont {J.}~\bibnamefont
  {Zou}}, \bibinfo {author} {\bibfnamefont {R.}~\bibnamefont {Zhu}}, \bibinfo
  {author} {\bibfnamefont {J.}~\bibnamefont {Wang}}, \bibinfo {author}
  {\bibfnamefont {H.}~\bibnamefont {Meng}}, \bibinfo {author} {\bibfnamefont
  {Z.}~\bibnamefont {Wang}}, \bibinfo {author} {\bibfnamefont {H.}~\bibnamefont
  {Chen}},\ and\ \bibinfo {author} {\bibfnamefont {Y.-X.}\ \bibnamefont
  {Weng}},\ }\bibfield  {title} {\bibinfo {title} {Coherent phonon-mediated
  many-body interaction in monolayer {WSe$_2$}},\ }\href
  {https://doi.org/10.1021/acs.jpclett.3c00870} {\bibfield  {journal} {\bibinfo
   {journal} {The Journal of Physical Chemistry Letters}\ }\textbf {\bibinfo
  {volume} {14}},\ \bibinfo {pages} {4657} (\bibinfo {year}
  {2023})}\BibitemShut {NoStop}%
\bibitem [{\citenamefont {Timmer}\ \emph {et~al.}(2024)\citenamefont {Timmer},
  \citenamefont {Gittinger}, \citenamefont {Quenzel}, \citenamefont {Cadore},
  \citenamefont {Rosa}, \citenamefont {Li}, \citenamefont {Soavi},
  \citenamefont {Lunemann}, \citenamefont {Stephan}, \citenamefont {Silies}
  \emph {et~al.}}]{timmer2024ultrafast}%
  \BibitemOpen
  \bibfield  {author} {\bibinfo {author} {\bibfnamefont {D.}~\bibnamefont
  {Timmer}}, \bibinfo {author} {\bibfnamefont {M.}~\bibnamefont {Gittinger}},
  \bibinfo {author} {\bibfnamefont {T.}~\bibnamefont {Quenzel}}, \bibinfo
  {author} {\bibfnamefont {A.~R.}\ \bibnamefont {Cadore}}, \bibinfo {author}
  {\bibfnamefont {B.~L.}\ \bibnamefont {Rosa}}, \bibinfo {author}
  {\bibfnamefont {W.}~\bibnamefont {Li}}, \bibinfo {author} {\bibfnamefont
  {G.}~\bibnamefont {Soavi}}, \bibinfo {author} {\bibfnamefont {D.~C.}\
  \bibnamefont {Lunemann}}, \bibinfo {author} {\bibfnamefont {S.}~\bibnamefont
  {Stephan}}, \bibinfo {author} {\bibfnamefont {M.}~\bibnamefont {Silies}},
  \emph {et~al.},\ }\bibfield  {title} {\bibinfo {title} {Ultrafast coherent
  exciton couplings and many-body interactions in monolayer {WS$_2$}},\ }\href
  {https://doi.org/10.1021/acs.nanolett.4c01991} {\bibfield  {journal}
  {\bibinfo  {journal} {Nano Letters}\ }\textbf {\bibinfo {volume} {24}},\
  \bibinfo {pages} {8117} (\bibinfo {year} {2024})}\BibitemShut {NoStop}%
\bibitem [{\citenamefont {Lloyd}\ \emph {et~al.}(2021)\citenamefont {Lloyd},
  \citenamefont {Wood}, \citenamefont {Mujid}, \citenamefont {Sohoni},
  \citenamefont {Ji}, \citenamefont {Ting}, \citenamefont {Higgins},
  \citenamefont {Park},\ and\ \citenamefont {Engel}}]{lloyd2021sub}%
  \BibitemOpen
  \bibfield  {author} {\bibinfo {author} {\bibfnamefont {L.~T.}\ \bibnamefont
  {Lloyd}}, \bibinfo {author} {\bibfnamefont {R.~E.}\ \bibnamefont {Wood}},
  \bibinfo {author} {\bibfnamefont {F.}~\bibnamefont {Mujid}}, \bibinfo
  {author} {\bibfnamefont {S.}~\bibnamefont {Sohoni}}, \bibinfo {author}
  {\bibfnamefont {K.~L.}\ \bibnamefont {Ji}}, \bibinfo {author} {\bibfnamefont
  {P.-C.}\ \bibnamefont {Ting}}, \bibinfo {author} {\bibfnamefont {J.~S.}\
  \bibnamefont {Higgins}}, \bibinfo {author} {\bibfnamefont {J.}~\bibnamefont
  {Park}},\ and\ \bibinfo {author} {\bibfnamefont {G.~S.}\ \bibnamefont
  {Engel}},\ }\bibfield  {title} {\bibinfo {title} {Sub-10 fs intervalley
  exciton coupling in monolayer {MoS$_2$} revealed by helicity-resolved
  two-dimensional electronic spectroscopy},\ }\href
  {https://doi.org/10.1021/acsnano.1c02381} {\bibfield  {journal} {\bibinfo
  {journal} {ACS Nano}\ }\textbf {\bibinfo {volume} {15}},\ \bibinfo {pages}
  {10253} (\bibinfo {year} {2021})}\BibitemShut {NoStop}%
\bibitem [{\citenamefont {Manca}\ \emph {et~al.}(2017)\citenamefont {Manca},
  \citenamefont {Glazov}, \citenamefont {Robert}, \citenamefont {Cadiz},
  \citenamefont {Taniguchi}, \citenamefont {Watanabe}, \citenamefont
  {Courtade}, \citenamefont {Amand}, \citenamefont {Renucci}, \citenamefont
  {Marie} \emph {et~al.}}]{manca2017enabling}%
  \BibitemOpen
  \bibfield  {author} {\bibinfo {author} {\bibfnamefont {M.}~\bibnamefont
  {Manca}}, \bibinfo {author} {\bibfnamefont {M.~M.}\ \bibnamefont {Glazov}},
  \bibinfo {author} {\bibfnamefont {C.}~\bibnamefont {Robert}}, \bibinfo
  {author} {\bibfnamefont {F.}~\bibnamefont {Cadiz}}, \bibinfo {author}
  {\bibfnamefont {T.}~\bibnamefont {Taniguchi}}, \bibinfo {author}
  {\bibfnamefont {K.}~\bibnamefont {Watanabe}}, \bibinfo {author}
  {\bibfnamefont {E.}~\bibnamefont {Courtade}}, \bibinfo {author}
  {\bibfnamefont {T.}~\bibnamefont {Amand}}, \bibinfo {author} {\bibfnamefont
  {P.}~\bibnamefont {Renucci}}, \bibinfo {author} {\bibfnamefont
  {X.}~\bibnamefont {Marie}}, \emph {et~al.},\ }\bibfield  {title} {\bibinfo
  {title} {Enabling valley selective exciton scattering in monolayer {WSe$_2$}
  through upconversion},\ }\href {https://doi.org/10.1038/ncomms14927}
  {\bibfield  {journal} {\bibinfo  {journal} {Nature Communications}\ }\textbf
  {\bibinfo {volume} {8}},\ \bibinfo {pages} {14927} (\bibinfo {year}
  {2017})}\BibitemShut {NoStop}%
\bibitem [{\citenamefont {Katsch}\ \emph {et~al.}(2019)\citenamefont {Katsch},
  \citenamefont {Selig},\ and\ \citenamefont {Knorr}}]{katsch2019theory}%
  \BibitemOpen
  \bibfield  {author} {\bibinfo {author} {\bibfnamefont {F.}~\bibnamefont
  {Katsch}}, \bibinfo {author} {\bibfnamefont {M.}~\bibnamefont {Selig}},\ and\
  \bibinfo {author} {\bibfnamefont {A.}~\bibnamefont {Knorr}},\ }\bibfield
  {title} {\bibinfo {title} {Theory of coherent pump--probe spectroscopy in
  monolayer transition metal dichalcogenides},\ }\href
  {https://doi.org/10.1088/2053-1583/ab5407} {\bibfield  {journal} {\bibinfo
  {journal} {2D Materials}\ }\textbf {\bibinfo {volume} {7}},\ \bibinfo {pages}
  {015021} (\bibinfo {year} {2019})}\BibitemShut {NoStop}%
\bibitem [{\citenamefont {Dobryakov}\ \emph {et~al.}(2010)\citenamefont
  {Dobryakov}, \citenamefont {Kovalenko}, \citenamefont {Weigel}, \citenamefont
  {P{\'e}rez-Lustres}, \citenamefont {Lange}, \citenamefont {M{\"u}ller},\ and\
  \citenamefont {Ernsting}}]{dobryakov2010femtosecond}%
  \BibitemOpen
  \bibfield  {author} {\bibinfo {author} {\bibfnamefont {A.}~\bibnamefont
  {Dobryakov}}, \bibinfo {author} {\bibfnamefont {S.~A.}\ \bibnamefont
  {Kovalenko}}, \bibinfo {author} {\bibfnamefont {A.}~\bibnamefont {Weigel}},
  \bibinfo {author} {\bibfnamefont {J.~L.}\ \bibnamefont {P{\'e}rez-Lustres}},
  \bibinfo {author} {\bibfnamefont {J.}~\bibnamefont {Lange}}, \bibinfo
  {author} {\bibfnamefont {A.}~\bibnamefont {M{\"u}ller}},\ and\ \bibinfo
  {author} {\bibfnamefont {N.}~\bibnamefont {Ernsting}},\ }\bibfield  {title}
  {\bibinfo {title} {Femtosecond pump/supercontinuum-probe spectroscopy:
  Optimized setup and signal analysis for single-shot spectral referencing},\
  }\href {https://doi.org/10.1063/1.3492897} {\bibfield  {journal} {\bibinfo
  {journal} {Review of Scientific Instruments}\ }\textbf {\bibinfo {volume}
  {81}},\ \bibinfo {pages} {113106} (\bibinfo {year} {2010})}\BibitemShut
  {NoStop}%
\bibitem [{\citenamefont {Kovalenko}\ \emph {et~al.}(1999)\citenamefont
  {Kovalenko}, \citenamefont {Dobryakov}, \citenamefont {Ruthmann},\ and\
  \citenamefont {Ernsting}}]{kovalenko1999femtosecond}%
  \BibitemOpen
  \bibfield  {author} {\bibinfo {author} {\bibfnamefont {S.~A.}\ \bibnamefont
  {Kovalenko}}, \bibinfo {author} {\bibfnamefont {A.~L.}\ \bibnamefont
  {Dobryakov}}, \bibinfo {author} {\bibfnamefont {J.}~\bibnamefont
  {Ruthmann}},\ and\ \bibinfo {author} {\bibfnamefont {N.~P.}\ \bibnamefont
  {Ernsting}},\ }\bibfield  {title} {\bibinfo {title} {Femtosecond spectroscopy
  of condensed phases with chirped supercontinuum probing},\ }\href
  {https://doi.org/10.1103/PhysRevA.59.2369} {\bibfield  {journal} {\bibinfo
  {journal} {Physical Review A}\ }\textbf {\bibinfo {volume} {59}},\ \bibinfo
  {pages} {2369} (\bibinfo {year} {1999})}\BibitemShut {NoStop}%
\bibitem [{\citenamefont {Desai}\ \emph {et~al.}(2016)\citenamefont {Desai},
  \citenamefont {Madhvapathy}, \citenamefont {Amani}, \citenamefont {Kiriya},
  \citenamefont {Hettick}, \citenamefont {Tosun}, \citenamefont {Zhou},
  \citenamefont {Dubey}, \citenamefont {Ager~III}, \citenamefont {Chrzan} \emph
  {et~al.}}]{desai2016gold}%
  \BibitemOpen
  \bibfield  {author} {\bibinfo {author} {\bibfnamefont {S.~B.}\ \bibnamefont
  {Desai}}, \bibinfo {author} {\bibfnamefont {S.~R.}\ \bibnamefont
  {Madhvapathy}}, \bibinfo {author} {\bibfnamefont {M.}~\bibnamefont {Amani}},
  \bibinfo {author} {\bibfnamefont {D.}~\bibnamefont {Kiriya}}, \bibinfo
  {author} {\bibfnamefont {M.}~\bibnamefont {Hettick}}, \bibinfo {author}
  {\bibfnamefont {M.}~\bibnamefont {Tosun}}, \bibinfo {author} {\bibfnamefont
  {Y.}~\bibnamefont {Zhou}}, \bibinfo {author} {\bibfnamefont {M.}~\bibnamefont
  {Dubey}}, \bibinfo {author} {\bibfnamefont {J.~W.}\ \bibnamefont {Ager~III}},
  \bibinfo {author} {\bibfnamefont {D.}~\bibnamefont {Chrzan}}, \emph
  {et~al.},\ }\bibfield  {title} {\bibinfo {title} {Gold-mediated exfoliation
  of ultralarge optoelectronically-perfect monolayers},\ }\href
  {https://doi.org/10.1002/adma.201506171} {\bibfield  {journal} {\bibinfo
  {journal} {Advanced Materials}\ }\textbf {\bibinfo {volume} {28}},\ \bibinfo
  {pages} {4053} (\bibinfo {year} {2016})}\BibitemShut {NoStop}%
\bibitem [{\citenamefont {Liu}\ \emph {et~al.}(2019)\citenamefont {Liu},
  \citenamefont {Ziffer}, \citenamefont {Hansen}, \citenamefont {Wang},\ and\
  \citenamefont {Zhu}}]{liu2019direct}%
  \BibitemOpen
  \bibfield  {author} {\bibinfo {author} {\bibfnamefont {F.}~\bibnamefont
  {Liu}}, \bibinfo {author} {\bibfnamefont {M.~E.}\ \bibnamefont {Ziffer}},
  \bibinfo {author} {\bibfnamefont {K.~R.}\ \bibnamefont {Hansen}}, \bibinfo
  {author} {\bibfnamefont {J.}~\bibnamefont {Wang}},\ and\ \bibinfo {author}
  {\bibfnamefont {X.}~\bibnamefont {Zhu}},\ }\bibfield  {title} {\bibinfo
  {title} {Direct determination of band-gap renormalization in the photoexcited
  monolayer {MoS$_2$}},\ }\href
  {https://doi.org/10.1103/PhysRevLett.122.246803} {\bibfield  {journal}
  {\bibinfo  {journal} {Physical Review Letters}\ }\textbf {\bibinfo {volume}
  {122}},\ \bibinfo {pages} {246803} (\bibinfo {year} {2019})}\BibitemShut
  {NoStop}%
\bibitem [{\citenamefont {Olsen}\ \emph {et~al.}(2025)\citenamefont {Olsen},
  \citenamefont {Yoon}, \citenamefont {Holbrook}, \citenamefont {Thinel},
  \citenamefont {Holtzman}, \citenamefont {Liu}, \citenamefont {Hsieh},
  \citenamefont {Li}, \citenamefont {Xu}, \citenamefont {Rojas-Gatjens} \emph
  {et~al.}}]{olsen2025macroscopic}%
  \BibitemOpen
  \bibfield  {author} {\bibinfo {author} {\bibfnamefont {N.}~\bibnamefont
  {Olsen}}, \bibinfo {author} {\bibfnamefont {S.}~\bibnamefont {Yoon}},
  \bibinfo {author} {\bibfnamefont {M.}~\bibnamefont {Holbrook}}, \bibinfo
  {author} {\bibfnamefont {M.}~\bibnamefont {Thinel}}, \bibinfo {author}
  {\bibfnamefont {L.~N.}\ \bibnamefont {Holtzman}}, \bibinfo {author}
  {\bibfnamefont {Y.}~\bibnamefont {Liu}}, \bibinfo {author} {\bibfnamefont
  {V.}~\bibnamefont {Hsieh}}, \bibinfo {author} {\bibfnamefont
  {Y.}~\bibnamefont {Li}}, \bibinfo {author} {\bibfnamefont {D.~D.}\
  \bibnamefont {Xu}}, \bibinfo {author} {\bibfnamefont {E.}~\bibnamefont
  {Rojas-Gatjens}}, \emph {et~al.},\ }\bibfield  {title} {\bibinfo {title}
  {Macroscopic transition metal dichalcogenide monolayers from gold-tape
  exfoliation retain intrinsic properties},\ }\href
  {https://doi.org/10.1021/acs.nanolett.5c03371} {\bibfield  {journal}
  {\bibinfo  {journal} {Nano Letters}\ }\textbf {\bibinfo {volume} {25}},\
  \bibinfo {pages} {15198} (\bibinfo {year} {2025})}\BibitemShut {NoStop}%
\bibitem [{\citenamefont {Wang}\ \emph {et~al.}(2018)\citenamefont {Wang},
  \citenamefont {Chernikov}, \citenamefont {Glazov}, \citenamefont {Heinz},
  \citenamefont {Marie}, \citenamefont {Amand},\ and\ \citenamefont
  {Urbaszek}}]{wang2018colloquium}%
  \BibitemOpen
  \bibfield  {author} {\bibinfo {author} {\bibfnamefont {G.}~\bibnamefont
  {Wang}}, \bibinfo {author} {\bibfnamefont {A.}~\bibnamefont {Chernikov}},
  \bibinfo {author} {\bibfnamefont {M.~M.}\ \bibnamefont {Glazov}}, \bibinfo
  {author} {\bibfnamefont {T.~F.}\ \bibnamefont {Heinz}}, \bibinfo {author}
  {\bibfnamefont {X.}~\bibnamefont {Marie}}, \bibinfo {author} {\bibfnamefont
  {T.}~\bibnamefont {Amand}},\ and\ \bibinfo {author} {\bibfnamefont
  {B.}~\bibnamefont {Urbaszek}},\ }\bibfield  {title} {\bibinfo {title}
  {Colloquium: Excitons in atomically thin transition metal dichalcogenides},\
  }\href {https://doi.org/10.1103/RevModPhys.90.021001} {\bibfield  {journal}
  {\bibinfo  {journal} {Reviews of Modern Physics}\ }\textbf {\bibinfo {volume}
  {90}},\ \bibinfo {pages} {021001} (\bibinfo {year} {2018})}\BibitemShut
  {NoStop}%
\bibitem [{\citenamefont {Tanda~Bonkano}\ \emph {et~al.}(2024)\citenamefont
  {Tanda~Bonkano}, \citenamefont {Palato}, \citenamefont {Krumland},
  \citenamefont {Kovalenko}, \citenamefont {Schwendke}, \citenamefont
  {Guerrini}, \citenamefont {Li}, \citenamefont {Zhu}, \citenamefont {Cocchi},\
  and\ \citenamefont {St{\"a}hler}}]{tanda2024evidence}%
  \BibitemOpen
  \bibfield  {author} {\bibinfo {author} {\bibfnamefont {B.}~\bibnamefont
  {Tanda~Bonkano}}, \bibinfo {author} {\bibfnamefont {S.}~\bibnamefont
  {Palato}}, \bibinfo {author} {\bibfnamefont {J.}~\bibnamefont {Krumland}},
  \bibinfo {author} {\bibfnamefont {S.}~\bibnamefont {Kovalenko}}, \bibinfo
  {author} {\bibfnamefont {P.}~\bibnamefont {Schwendke}}, \bibinfo {author}
  {\bibfnamefont {M.}~\bibnamefont {Guerrini}}, \bibinfo {author}
  {\bibfnamefont {Q.}~\bibnamefont {Li}}, \bibinfo {author} {\bibfnamefont
  {X.}~\bibnamefont {Zhu}}, \bibinfo {author} {\bibfnamefont {C.}~\bibnamefont
  {Cocchi}},\ and\ \bibinfo {author} {\bibfnamefont {J.}~\bibnamefont
  {St{\"a}hler}},\ }\bibfield  {title} {\bibinfo {title} {Evidence for hybrid
  inorganic--organic transitions at the {WS$_2$}/terrylene interface},\ }\href
  {https://doi.org/10.1002/pssa.202300346} {\bibfield  {journal} {\bibinfo
  {journal} {Physica Status Solidi (a)}\ }\textbf {\bibinfo {volume} {221}},\
  \bibinfo {pages} {2300346} (\bibinfo {year} {2024})}\BibitemShut {NoStop}%
\bibitem [{\citenamefont {Li}\ and\ \citenamefont {Heinz}(2018)}]{li2018two}%
  \BibitemOpen
  \bibfield  {author} {\bibinfo {author} {\bibfnamefont {Y.}~\bibnamefont
  {Li}}\ and\ \bibinfo {author} {\bibfnamefont {T.~F.}\ \bibnamefont {Heinz}},\
  }\bibfield  {title} {\bibinfo {title} {Two-dimensional models for the optical
  response of thin films},\ }\href {https://doi.org/10.1088/2053-1583/aab0cf}
  {\bibfield  {journal} {\bibinfo  {journal} {2D Materials}\ }\textbf {\bibinfo
  {volume} {5}},\ \bibinfo {pages} {025021} (\bibinfo {year}
  {2018})}\BibitemShut {NoStop}%
\bibitem [{\citenamefont {Guo}\ \emph {et~al.}(2019)\citenamefont {Guo},
  \citenamefont {Wu}, \citenamefont {Cao}, \citenamefont {Monahan},
  \citenamefont {Lee}, \citenamefont {Louie},\ and\ \citenamefont
  {Fleming}}]{guo2019exchange}%
  \BibitemOpen
  \bibfield  {author} {\bibinfo {author} {\bibfnamefont {L.}~\bibnamefont
  {Guo}}, \bibinfo {author} {\bibfnamefont {M.}~\bibnamefont {Wu}}, \bibinfo
  {author} {\bibfnamefont {T.}~\bibnamefont {Cao}}, \bibinfo {author}
  {\bibfnamefont {D.~M.}\ \bibnamefont {Monahan}}, \bibinfo {author}
  {\bibfnamefont {Y.-H.}\ \bibnamefont {Lee}}, \bibinfo {author} {\bibfnamefont
  {S.~G.}\ \bibnamefont {Louie}},\ and\ \bibinfo {author} {\bibfnamefont
  {G.~R.}\ \bibnamefont {Fleming}},\ }\bibfield  {title} {\bibinfo {title}
  {Exchange-driven intravalley mixing of excitons in monolayer transition metal
  dichalcogenides},\ }\href {https://doi.org/10.1038/s41567-018-0362-y}
  {\bibfield  {journal} {\bibinfo  {journal} {Nature Physics}\ }\textbf
  {\bibinfo {volume} {15}},\ \bibinfo {pages} {228} (\bibinfo {year}
  {2019})}\BibitemShut {NoStop}%
\bibitem [{\citenamefont {Bange}\ \emph {et~al.}(2023)\citenamefont {Bange},
  \citenamefont {Werner}, \citenamefont {Schmitt}, \citenamefont {Bennecke},
  \citenamefont {Meneghini}, \citenamefont {AlMutairi}, \citenamefont
  {Merboldt}, \citenamefont {Watanabe}, \citenamefont {Taniguchi},
  \citenamefont {Steil} \emph {et~al.}}]{bange2023ultrafast}%
  \BibitemOpen
  \bibfield  {author} {\bibinfo {author} {\bibfnamefont {J.~P.}\ \bibnamefont
  {Bange}}, \bibinfo {author} {\bibfnamefont {P.}~\bibnamefont {Werner}},
  \bibinfo {author} {\bibfnamefont {D.}~\bibnamefont {Schmitt}}, \bibinfo
  {author} {\bibfnamefont {W.}~\bibnamefont {Bennecke}}, \bibinfo {author}
  {\bibfnamefont {G.}~\bibnamefont {Meneghini}}, \bibinfo {author}
  {\bibfnamefont {A.}~\bibnamefont {AlMutairi}}, \bibinfo {author}
  {\bibfnamefont {M.}~\bibnamefont {Merboldt}}, \bibinfo {author}
  {\bibfnamefont {K.}~\bibnamefont {Watanabe}}, \bibinfo {author}
  {\bibfnamefont {T.}~\bibnamefont {Taniguchi}}, \bibinfo {author}
  {\bibfnamefont {S.}~\bibnamefont {Steil}}, \emph {et~al.},\ }\bibfield
  {title} {\bibinfo {title} {Ultrafast dynamics of bright and dark excitons in
  monolayer {WSe$_2$} and heterobilayer {WSe$_2$}/{MoS$_2$}},\ }\href
  {https://doi.org/10.1088/2053-1583/ace067} {\bibfield  {journal} {\bibinfo
  {journal} {2D Materials}\ }\textbf {\bibinfo {volume} {10}},\ \bibinfo
  {pages} {035039} (\bibinfo {year} {2023})}\BibitemShut {NoStop}%
\bibitem [{\citenamefont {Wang}\ \emph {et~al.}(2024)\citenamefont {Wang},
  \citenamefont {Nishida}, \citenamefont {Nakamoto}, \citenamefont {Yang},
  \citenamefont {Sakuma}, \citenamefont {Zhang}, \citenamefont {Endo},
  \citenamefont {Miyata},\ and\ \citenamefont {Kumagai}}]{wang2024ultrafast}%
  \BibitemOpen
  \bibfield  {author} {\bibinfo {author} {\bibfnamefont {Y.}~\bibnamefont
  {Wang}}, \bibinfo {author} {\bibfnamefont {J.}~\bibnamefont {Nishida}},
  \bibinfo {author} {\bibfnamefont {K.}~\bibnamefont {Nakamoto}}, \bibinfo
  {author} {\bibfnamefont {X.}~\bibnamefont {Yang}}, \bibinfo {author}
  {\bibfnamefont {Y.}~\bibnamefont {Sakuma}}, \bibinfo {author} {\bibfnamefont
  {W.}~\bibnamefont {Zhang}}, \bibinfo {author} {\bibfnamefont
  {T.}~\bibnamefont {Endo}}, \bibinfo {author} {\bibfnamefont {Y.}~\bibnamefont
  {Miyata}},\ and\ \bibinfo {author} {\bibfnamefont {T.}~\bibnamefont
  {Kumagai}},\ }\bibfield  {title} {\bibinfo {title} {Ultrafast nano-imaging of
  spatially modulated many-body dynamics in {CVD}-grown monolayer {WS$_2$}},\
  }\href {https://doi.org/10.1021/acsphotonics.4c01545} {\bibfield  {journal}
  {\bibinfo  {journal} {ACS Photonics}\ }\textbf {\bibinfo {volume} {12}},\
  \bibinfo {pages} {207} (\bibinfo {year} {2024})}\BibitemShut {NoStop}%
\bibitem [{\citenamefont {Selig}\ \emph {et~al.}(2016)\citenamefont {Selig},
  \citenamefont {Bergh{\"a}user}, \citenamefont {Raja}, \citenamefont {Nagler},
  \citenamefont {Sch{\"u}ller}, \citenamefont {Heinz}, \citenamefont {Korn},
  \citenamefont {Chernikov}, \citenamefont {Malic},\ and\ \citenamefont
  {Knorr}}]{selig2016excitonic}%
  \BibitemOpen
  \bibfield  {author} {\bibinfo {author} {\bibfnamefont {M.}~\bibnamefont
  {Selig}}, \bibinfo {author} {\bibfnamefont {G.}~\bibnamefont
  {Bergh{\"a}user}}, \bibinfo {author} {\bibfnamefont {A.}~\bibnamefont
  {Raja}}, \bibinfo {author} {\bibfnamefont {P.}~\bibnamefont {Nagler}},
  \bibinfo {author} {\bibfnamefont {C.}~\bibnamefont {Sch{\"u}ller}}, \bibinfo
  {author} {\bibfnamefont {T.~F.}\ \bibnamefont {Heinz}}, \bibinfo {author}
  {\bibfnamefont {T.}~\bibnamefont {Korn}}, \bibinfo {author} {\bibfnamefont
  {A.}~\bibnamefont {Chernikov}}, \bibinfo {author} {\bibfnamefont
  {E.}~\bibnamefont {Malic}},\ and\ \bibinfo {author} {\bibfnamefont
  {A.}~\bibnamefont {Knorr}},\ }\bibfield  {title} {\bibinfo {title} {Excitonic
  linewidth and coherence lifetime in monolayer transition metal
  dichalcogenides},\ }\href {https://doi.org/10.1038/ncomms13279} {\bibfield
  {journal} {\bibinfo  {journal} {Nature Communications}\ }\textbf {\bibinfo
  {volume} {7}},\ \bibinfo {pages} {13279} (\bibinfo {year}
  {2016})}\BibitemShut {NoStop}%
\bibitem [{\citenamefont {Sim}\ \emph {et~al.}(2013)\citenamefont {Sim},
  \citenamefont {Park}, \citenamefont {Song}, \citenamefont {In}, \citenamefont
  {Lee}, \citenamefont {Kim},\ and\ \citenamefont {Choi}}]{sim2013exciton}%
  \BibitemOpen
  \bibfield  {author} {\bibinfo {author} {\bibfnamefont {S.}~\bibnamefont
  {Sim}}, \bibinfo {author} {\bibfnamefont {J.}~\bibnamefont {Park}}, \bibinfo
  {author} {\bibfnamefont {J.-G.}\ \bibnamefont {Song}}, \bibinfo {author}
  {\bibfnamefont {C.}~\bibnamefont {In}}, \bibinfo {author} {\bibfnamefont
  {Y.-S.}\ \bibnamefont {Lee}}, \bibinfo {author} {\bibfnamefont
  {H.}~\bibnamefont {Kim}},\ and\ \bibinfo {author} {\bibfnamefont
  {H.}~\bibnamefont {Choi}},\ }\bibfield  {title} {\bibinfo {title} {Exciton
  dynamics in atomically thin {MoS$_2$}: interexcitonic interaction and
  broadening kinetics},\ }\href {https://doi.org/10.1103/PhysRevB.88.075434}
  {\bibfield  {journal} {\bibinfo  {journal} {Physical Review B}\ }\textbf
  {\bibinfo {volume} {88}},\ \bibinfo {pages} {075434} (\bibinfo {year}
  {2013})}\BibitemShut {NoStop}%
\bibitem [{\citenamefont {Gupta}\ and\ \citenamefont
  {Majumdar}(2019)}]{gupta2019fundamental}%
  \BibitemOpen
  \bibfield  {author} {\bibinfo {author} {\bibfnamefont {G.}~\bibnamefont
  {Gupta}}\ and\ \bibinfo {author} {\bibfnamefont {K.}~\bibnamefont
  {Majumdar}},\ }\bibfield  {title} {\bibinfo {title} {Fundamental exciton
  linewidth broadening in monolayer transition metal dichalcogenides},\ }\href
  {https://doi.org/10.1103/PhysRevB.99.085412} {\bibfield  {journal} {\bibinfo
  {journal} {Physical Review B}\ }\textbf {\bibinfo {volume} {99}},\ \bibinfo
  {pages} {085412} (\bibinfo {year} {2019})}\BibitemShut {NoStop}%
\bibitem [{\citenamefont {Moody}\ \emph {et~al.}(2015)\citenamefont {Moody},
  \citenamefont {Kavir~Dass}, \citenamefont {Hao}, \citenamefont {Chen},
  \citenamefont {Li}, \citenamefont {Singh}, \citenamefont {Tran},
  \citenamefont {Clark}, \citenamefont {Xu}, \citenamefont {Bergh{\"a}user}
  \emph {et~al.}}]{moody2015intrinsic}%
  \BibitemOpen
  \bibfield  {author} {\bibinfo {author} {\bibfnamefont {G.}~\bibnamefont
  {Moody}}, \bibinfo {author} {\bibfnamefont {C.}~\bibnamefont {Kavir~Dass}},
  \bibinfo {author} {\bibfnamefont {K.}~\bibnamefont {Hao}}, \bibinfo {author}
  {\bibfnamefont {C.-H.}\ \bibnamefont {Chen}}, \bibinfo {author}
  {\bibfnamefont {L.-J.}\ \bibnamefont {Li}}, \bibinfo {author} {\bibfnamefont
  {A.}~\bibnamefont {Singh}}, \bibinfo {author} {\bibfnamefont
  {K.}~\bibnamefont {Tran}}, \bibinfo {author} {\bibfnamefont {G.}~\bibnamefont
  {Clark}}, \bibinfo {author} {\bibfnamefont {X.}~\bibnamefont {Xu}}, \bibinfo
  {author} {\bibfnamefont {G.}~\bibnamefont {Bergh{\"a}user}}, \emph {et~al.},\
  }\bibfield  {title} {\bibinfo {title} {Intrinsic homogeneous linewidth and
  broadening mechanisms of excitons in monolayer transition metal
  dichalcogenides},\ }\href {https://doi.org/10.1038/ncomms9315} {\bibfield
  {journal} {\bibinfo  {journal} {Nature Communications}\ }\textbf {\bibinfo
  {volume} {6}},\ \bibinfo {pages} {8315} (\bibinfo {year} {2015})}\BibitemShut
  {NoStop}%
\bibitem [{\citenamefont {Mad{\'e}o}\ \emph {et~al.}(2020)\citenamefont
  {Mad{\'e}o}, \citenamefont {Man}, \citenamefont {Sahoo}, \citenamefont
  {Campbell}, \citenamefont {Pareek}, \citenamefont {Wong}, \citenamefont
  {Al-Mahboob}, \citenamefont {Chan}, \citenamefont {Karmakar}, \citenamefont
  {Mariserla} \emph {et~al.}}]{madeo2020directly}%
  \BibitemOpen
  \bibfield  {author} {\bibinfo {author} {\bibfnamefont {J.}~\bibnamefont
  {Mad{\'e}o}}, \bibinfo {author} {\bibfnamefont {M.~K.}\ \bibnamefont {Man}},
  \bibinfo {author} {\bibfnamefont {C.}~\bibnamefont {Sahoo}}, \bibinfo
  {author} {\bibfnamefont {M.}~\bibnamefont {Campbell}}, \bibinfo {author}
  {\bibfnamefont {V.}~\bibnamefont {Pareek}}, \bibinfo {author} {\bibfnamefont
  {E.~L.}\ \bibnamefont {Wong}}, \bibinfo {author} {\bibfnamefont
  {A.}~\bibnamefont {Al-Mahboob}}, \bibinfo {author} {\bibfnamefont {N.~S.}\
  \bibnamefont {Chan}}, \bibinfo {author} {\bibfnamefont {A.}~\bibnamefont
  {Karmakar}}, \bibinfo {author} {\bibfnamefont {B.~M.~K.}\ \bibnamefont
  {Mariserla}}, \emph {et~al.},\ }\bibfield  {title} {\bibinfo {title}
  {Directly visualizing the momentum-forbidden dark excitons and their dynamics
  in atomically thin semiconductors},\ }\href
  {https://doi.org/10.1126/science.aba1029} {\bibfield  {journal} {\bibinfo
  {journal} {Science}\ }\textbf {\bibinfo {volume} {370}},\ \bibinfo {pages}
  {1199} (\bibinfo {year} {2020})}\BibitemShut {NoStop}%
\bibitem [{\citenamefont {Wallauer}\ \emph {et~al.}(2021)\citenamefont
  {Wallauer}, \citenamefont {Perea-Causin}, \citenamefont {M{\"u}nster},
  \citenamefont {Zajusch}, \citenamefont {Brem}, \citenamefont {G{\"u}dde},
  \citenamefont {Tanimura}, \citenamefont {Lin}, \citenamefont {Huber},
  \citenamefont {Malic} \emph {et~al.}}]{wallauer2021momentum}%
  \BibitemOpen
  \bibfield  {author} {\bibinfo {author} {\bibfnamefont {R.}~\bibnamefont
  {Wallauer}}, \bibinfo {author} {\bibfnamefont {R.}~\bibnamefont
  {Perea-Causin}}, \bibinfo {author} {\bibfnamefont {L.}~\bibnamefont
  {M{\"u}nster}}, \bibinfo {author} {\bibfnamefont {S.}~\bibnamefont
  {Zajusch}}, \bibinfo {author} {\bibfnamefont {S.}~\bibnamefont {Brem}},
  \bibinfo {author} {\bibfnamefont {J.}~\bibnamefont {G{\"u}dde}}, \bibinfo
  {author} {\bibfnamefont {K.}~\bibnamefont {Tanimura}}, \bibinfo {author}
  {\bibfnamefont {K.-Q.}\ \bibnamefont {Lin}}, \bibinfo {author} {\bibfnamefont
  {R.}~\bibnamefont {Huber}}, \bibinfo {author} {\bibfnamefont
  {E.}~\bibnamefont {Malic}}, \emph {et~al.},\ }\bibfield  {title} {\bibinfo
  {title} {Momentum-resolved observation of exciton formation dynamics in
  monolayer {WS$_2$}},\ }\href {https://doi.org/10.1021/acs.nanolett.1c01839}
  {\bibfield  {journal} {\bibinfo  {journal} {Nano Letters}\ }\textbf {\bibinfo
  {volume} {21}},\ \bibinfo {pages} {5867} (\bibinfo {year}
  {2021})}\BibitemShut {NoStop}%
\bibitem [{\citenamefont {Wang}\ \emph {et~al.}(2016)\citenamefont {Wang},
  \citenamefont {Zhang}, \citenamefont {Chan}, \citenamefont {Manolatou},
  \citenamefont {Tiwari},\ and\ \citenamefont {Rana}}]{wang2016radiative}%
  \BibitemOpen
  \bibfield  {author} {\bibinfo {author} {\bibfnamefont {H.}~\bibnamefont
  {Wang}}, \bibinfo {author} {\bibfnamefont {C.}~\bibnamefont {Zhang}},
  \bibinfo {author} {\bibfnamefont {W.}~\bibnamefont {Chan}}, \bibinfo {author}
  {\bibfnamefont {C.}~\bibnamefont {Manolatou}}, \bibinfo {author}
  {\bibfnamefont {S.}~\bibnamefont {Tiwari}},\ and\ \bibinfo {author}
  {\bibfnamefont {F.}~\bibnamefont {Rana}},\ }\bibfield  {title} {\bibinfo
  {title} {Radiative lifetimes of excitons and trions in monolayers of the
  metal dichalcogenide {MoS$_2$}},\ }\href
  {https://doi.org/10.1103/PhysRevB.93.045407} {\bibfield  {journal} {\bibinfo
  {journal} {Physical Review B}\ }\textbf {\bibinfo {volume} {93}},\ \bibinfo
  {pages} {045407} (\bibinfo {year} {2016})}\BibitemShut {NoStop}%
\bibitem [{\citenamefont {Palummo}\ \emph {et~al.}(2015)\citenamefont
  {Palummo}, \citenamefont {Bernardi},\ and\ \citenamefont
  {Grossman}}]{palummo2015exciton}%
  \BibitemOpen
  \bibfield  {author} {\bibinfo {author} {\bibfnamefont {M.}~\bibnamefont
  {Palummo}}, \bibinfo {author} {\bibfnamefont {M.}~\bibnamefont {Bernardi}},\
  and\ \bibinfo {author} {\bibfnamefont {J.~C.}\ \bibnamefont {Grossman}},\
  }\bibfield  {title} {\bibinfo {title} {Exciton radiative lifetimes in
  two-dimensional transition metal dichalcogenides},\ }\href
  {https://doi.org/10.1021/nl503799t} {\bibfield  {journal} {\bibinfo
  {journal} {Nano Letters}\ }\textbf {\bibinfo {volume} {15}},\ \bibinfo
  {pages} {2794} (\bibinfo {year} {2015})}\BibitemShut {NoStop}%
\bibitem [{\citenamefont {Erkensten}\ \emph {et~al.}(2021)\citenamefont
  {Erkensten}, \citenamefont {Brem},\ and\ \citenamefont
  {Malic}}]{Erkensten2021exciton}%
  \BibitemOpen
  \bibfield  {author} {\bibinfo {author} {\bibfnamefont {D.}~\bibnamefont
  {Erkensten}}, \bibinfo {author} {\bibfnamefont {S.}~\bibnamefont {Brem}},\
  and\ \bibinfo {author} {\bibfnamefont {E.}~\bibnamefont {Malic}},\ }\bibfield
   {title} {\bibinfo {title} {Exciton-exciton interaction in transition metal
  dichalcogenide monolayers and van der {W}aals heterostructures},\ }\href
  {https://doi.org/10.1103/PhysRevB.103.045426} {\bibfield  {journal} {\bibinfo
   {journal} {Physical Review B}\ }\textbf {\bibinfo {volume} {103}},\ \bibinfo
  {pages} {045426} (\bibinfo {year} {2021})}\BibitemShut {NoStop}%
\bibitem [{\citenamefont {Mohapatra}(2026)}]{mohapatra2026Photoinduced}%
  \BibitemOpen
  \bibfield  {author} {\bibinfo {author} {\bibfnamefont {S.}~\bibnamefont
  {Mohapatra}},\ }\emph {\bibinfo {title} {Photoinduced Ultrafast and
  Discontinuous Exciton Mott Transition in Monolayer {WS$_2$}}},\ \href
  {https://doi.org/10.18452/38546} {Ph.D. thesis},\ \bibinfo  {school}
  {Humboldt University of Berlin} (\bibinfo {year} {2026})\BibitemShut
  {NoStop}%
\bibitem [{Note1()}]{Note1}%
  \BibitemOpen
  \bibinfo {note} {It should be noted that above the critical excitation
  density of 31~x~10$^{12}$cm$^{-2}$, this component seems to be missing. This
  likely just because it is accelerated and does not appear at 200\ fs anymore
  (cf. supplementary information section S6)}\BibitemShut {NoStop}%
\bibitem [{\citenamefont {Champagne}\ \emph {et~al.}(2023)\citenamefont
  {Champagne}, \citenamefont {Haber}, \citenamefont {Pokawanvit}, \citenamefont
  {Qiu}, \citenamefont {Biswas}, \citenamefont {Atwater}, \citenamefont
  {da~Jornada},\ and\ \citenamefont {Neaton}}]{champagne2023quasiparticle}%
  \BibitemOpen
  \bibfield  {author} {\bibinfo {author} {\bibfnamefont {A.}~\bibnamefont
  {Champagne}}, \bibinfo {author} {\bibfnamefont {J.~B.}\ \bibnamefont
  {Haber}}, \bibinfo {author} {\bibfnamefont {S.}~\bibnamefont {Pokawanvit}},
  \bibinfo {author} {\bibfnamefont {D.~Y.}\ \bibnamefont {Qiu}}, \bibinfo
  {author} {\bibfnamefont {S.}~\bibnamefont {Biswas}}, \bibinfo {author}
  {\bibfnamefont {H.~A.}\ \bibnamefont {Atwater}}, \bibinfo {author}
  {\bibfnamefont {F.~H.}\ \bibnamefont {da~Jornada}},\ and\ \bibinfo {author}
  {\bibfnamefont {J.~B.}\ \bibnamefont {Neaton}},\ }\bibfield  {title}
  {\bibinfo {title} {Quasiparticle and optical properties of carrier-doped
  monolayer {MoTe$_2$} from first principles},\ }\href
  {https://doi.org/10.1021/acs.nanolett.3c00386} {\bibfield  {journal}
  {\bibinfo  {journal} {Nano Letters}\ }\textbf {\bibinfo {volume} {23}},\
  \bibinfo {pages} {4274} (\bibinfo {year} {2023})}\BibitemShut {NoStop}%
\bibitem [{\citenamefont {Calati}\ \emph {et~al.}(2021)\citenamefont {Calati},
  \citenamefont {Li}, \citenamefont {Zhu},\ and\ \citenamefont
  {St{\"a}hler}}]{calati2021ultrafast}%
  \BibitemOpen
  \bibfield  {author} {\bibinfo {author} {\bibfnamefont {S.}~\bibnamefont
  {Calati}}, \bibinfo {author} {\bibfnamefont {Q.}~\bibnamefont {Li}}, \bibinfo
  {author} {\bibfnamefont {X.}~\bibnamefont {Zhu}},\ and\ \bibinfo {author}
  {\bibfnamefont {J.}~\bibnamefont {St{\"a}hler}},\ }\bibfield  {title}
  {\bibinfo {title} {Ultrafast evolution of the complex dielectric function of
  monolayer {WS$_2$} after photoexcitation},\ }\href
  {https://doi.org/10.1039/d1cp03437e} {\bibfield  {journal} {\bibinfo
  {journal} {Physical Chemistry Chemical Physics}\ }\textbf {\bibinfo {volume}
  {23}},\ \bibinfo {pages} {22640} (\bibinfo {year} {2021})}\BibitemShut
  {NoStop}%
\bibitem [{\citenamefont {Bauer}\ \emph {et~al.}(2015)\citenamefont {Bauer},
  \citenamefont {Marienfeld},\ and\ \citenamefont {Aeschlimann}}]{bauer2015}%
  \BibitemOpen
  \bibfield  {author} {\bibinfo {author} {\bibfnamefont {M.}~\bibnamefont
  {Bauer}}, \bibinfo {author} {\bibfnamefont {A.}~\bibnamefont {Marienfeld}},\
  and\ \bibinfo {author} {\bibfnamefont {M.}~\bibnamefont {Aeschlimann}},\
  }\bibfield  {title} {\bibinfo {title} {Hot electron lifetimes in metals
  probed by time-resolved two-photon photoemission},\ }\href
  {https://doi.org/10.1016/j.progsurf.2015.05.001} {\bibfield  {journal}
  {\bibinfo  {journal} {Progress in Surface Science}\ }\textbf {\bibinfo
  {volume} {90}},\ \bibinfo {pages} {319} (\bibinfo {year} {2015})}\BibitemShut
  {NoStop}%
\bibitem [{\citenamefont {Sekiguchi}\ \emph {et~al.}(2017)\citenamefont
  {Sekiguchi}, \citenamefont {Mochizuki}, \citenamefont {Kim}, \citenamefont
  {Akiyama}, \citenamefont {Pfeiffer}, \citenamefont {West},\ and\
  \citenamefont {Shimano}}]{sekiguchi2017anomalous}%
  \BibitemOpen
  \bibfield  {author} {\bibinfo {author} {\bibfnamefont {F.}~\bibnamefont
  {Sekiguchi}}, \bibinfo {author} {\bibfnamefont {T.}~\bibnamefont
  {Mochizuki}}, \bibinfo {author} {\bibfnamefont {C.}~\bibnamefont {Kim}},
  \bibinfo {author} {\bibfnamefont {H.}~\bibnamefont {Akiyama}}, \bibinfo
  {author} {\bibfnamefont {L.~N.}\ \bibnamefont {Pfeiffer}}, \bibinfo {author}
  {\bibfnamefont {K.~W.}\ \bibnamefont {West}},\ and\ \bibinfo {author}
  {\bibfnamefont {R.}~\bibnamefont {Shimano}},\ }\bibfield  {title} {\bibinfo
  {title} {Anomalous metal phase emergent on the verge of an exciton {Mott}
  transition},\ }\href {https://doi.org/10.1103/PhysRevLett.118.067401}
  {\bibfield  {journal} {\bibinfo  {journal} {Physical Review Letters}\
  }\textbf {\bibinfo {volume} {118}},\ \bibinfo {pages} {067401} (\bibinfo
  {year} {2017})}\BibitemShut {NoStop}%
\end{thebibliography}%




\section*{Supplementary Information}
\title{Supplementary material for: \\Discontinuous character of the ultrafast exciton Mott transition in monolayer WS$_2$}

\author{Subhadra Mohapatra}
\affiliation{Humboldt-Universität zu Berlin, Institut für Chemie, Berlin, Germany}

\author{Samuel Palato}
\affiliation{Humboldt-Universität zu Berlin, Institut für Chemie, Berlin, Germany}

\author{Nicholas Olsen}
\affiliation{Department of Chemistry, Columbia University, New York, New York 10027, USA}

\author{Julia Stähler}
\affiliation{Humboldt-Universität zu Berlin, Institut für Chemie, Berlin, Germany}
\affiliation{Fritz-Haber-Institut der Max-Planck-Gesellschaft, Abt. Physikalische Chemie, Berlin, Germany}
\author{Lukas Gierster}
\email{Corresponding author; lukas.gierster@hu-berlin.de}
\affiliation{Humboldt-Universität zu Berlin, Institut für Chemie, Berlin, Germany}



\maketitle
\setcounter{figure}{0}
\renewcommand{\thefigure}{S\arabic{figure}}
\setcounter{table}{0}
\renewcommand{\thetable}{S\arabic{table}}
\setcounter{section}{0}
\renewcommand{\thesection}{S\arabic{section}}

\section{Steady state absorption spectra}
\label{sec:S1}
\begin{figure}[h]
    \centering
    \includegraphics[width=8cm]{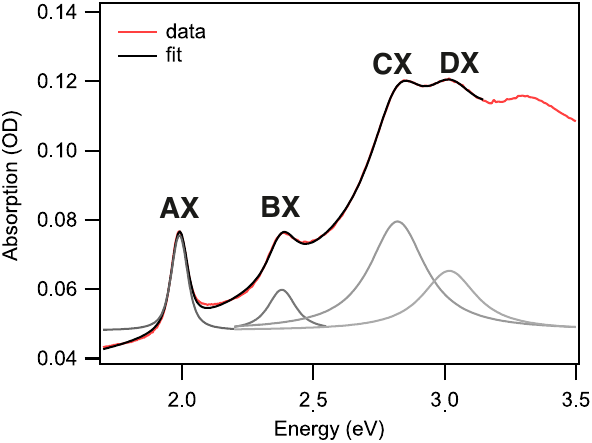}
    \caption{Steady-state absorption spectrum of the monolayer WS$_{2}$ on a fused-silica substrate, measured via UV-Vis spectroscopy, and the resulting fit based on Eq.~\ref{equation S1} as detailed in the text. The Voigt contribution of each excitonic resonances (AX, BX, CX, DX) extracted from the fit function with the offset ($y_0$), are plotted in grey lines. 
    }
    \label{fig: S1}
\end{figure}

Fig.~\ref{fig: S1} shows the steady state response of the monolayer WS$_2$. The peaks around 2 eV, 2.4 eV, 2.8 eV, and 3.0 eV correspond to the excitonic resonances (AX, BX, CX, DX) respectively, of the monolayer as reported previously in the literature \cite{cunningham2017photoinduced, calati2021ultrafast}. The steady state response is fitted using a sum of 4 Voigts (one corresponding to each excitonic resonance) plus a polynomial background:
 \begin{equation}
     f(E)= y_0 +\sum_{i=1}^4 V_i(S_i,\gamma_i,\sigma_i, E_i) + X(E-E_0)+ Y(E-E_0)^2 
     \label{equation S1}
 \end{equation}
where $y_0$ is the offset, $V_i$ for i=1 to 4, corresponds to each excitonic resonance (AX, BX, CX, and DX). The Voigt is defined by the parameters oscillator strength ($S_i$), Lorentzian width ($\gamma_i$), Gaussian width ($\sigma_i$) and resonance energy ($E_i$) corresponding to each excitonic resonance. X and Y are the coefficients of the polynomial with the center $E_0$ (corresponds to the resonance energy of AX; $E_0$ = 1.991 eV). The polynomial background disentangles the exciton responses from the continuous background absorption originating likely from the states with higher energy and from the underlying fused silica substrate. The fit function accurately describes the steady state absorption spectrum in the range from 1.8 eV to 3.2 eV. 

The fit resolved Voigt parameters of each excitonic resonance are detailed in the Table~\ref{tab:S1}. The Gaussian width ($\sigma$), which is kept common for all excitonic resonances, has a magnitude of 46 $\pm$ 2 meV. 

\begin{table}[h]
\small
\begin{tabular}{@{}cccc@{}} 
\toprule
Exciton Resonance  & $S$ (mOD$\cdot$eV) & $E$ (eV) & $\gamma$ (meV)\\
\midrule
AX & 2.7 $\pm$ 0.1 & 1.991 $\pm$ 0.001 & 44 $\pm$ 5 \\
BX &  2.2 $\pm$ 0.1 & 2.381 $\pm$ 0.001& 107 $\pm$ 11 \\
CX & 12.5 $\pm$ 0.3 & 2.821 $\pm$ 0.001& 247 $\pm$ 20\\
DX & 6.6 $\pm$ 0.7 & 3.018 $\pm$ 0.003 & 238 $\pm$ 22 \\

\bottomrule
\end{tabular}
\caption{Tabulates the fit-resolved Voigt parameters of all excitonic resonances (AX to DX) in the steady state absorption spectrum.}
\label{tab:S1}
\end{table}

\section{Comparison of 
${V}_0$ extracted from differential absorption spectra with the steady state response}

\begin{figure}
    \centering
    \includegraphics[width=0.92\linewidth]{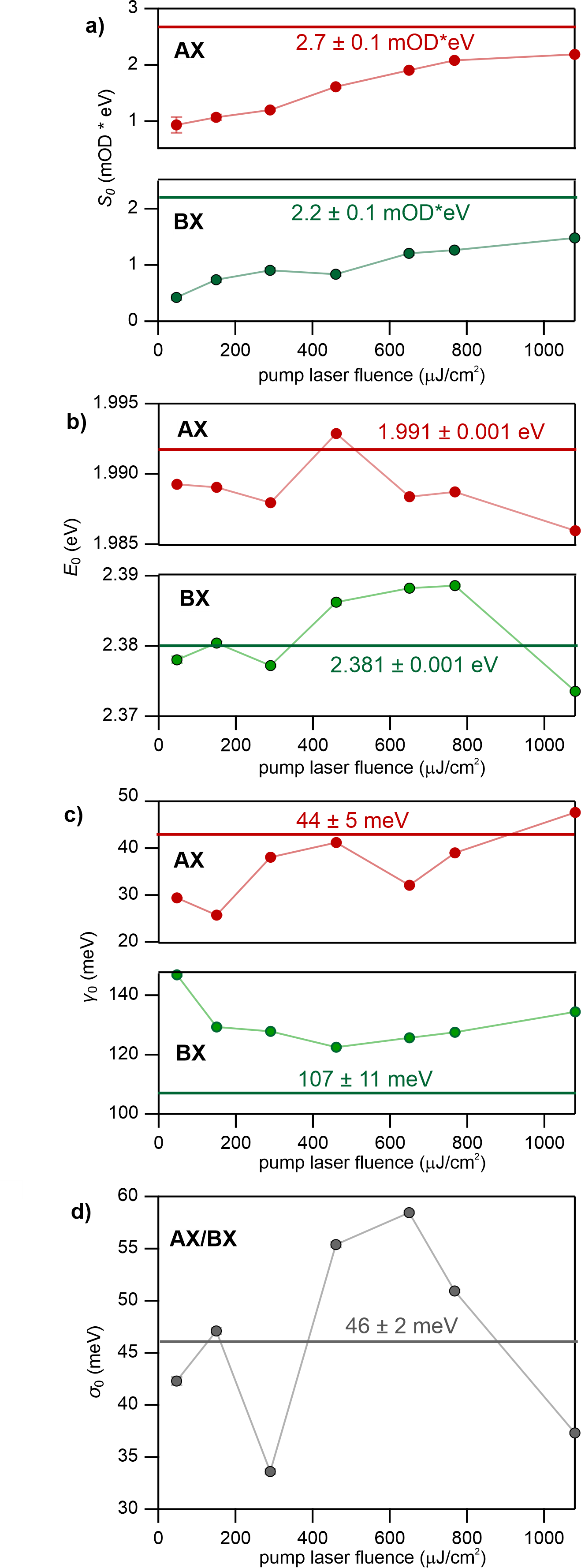}
    \caption{Global fit-retrieved $V_0$ parameters. (a) oscillator strength ($S_{0}$), (b) resonance energy ($E_{0}$), (c) Lorentzian width ($\gamma_{0}$) for both AX (upper panel) and BX (lower panel), respectively and (d) the Gaussian width ($\sigma_{0}$), common for both excitonic resonances as a function of pump laser fluence. The magnitude of the corresponding steady state parameter, retrieved from the fit in Figure \ref{fig: S1} are indicated as horizontal lines.}
    \label{fig:S2}
\end{figure}
This work employs a lineshape analysis model, where the differential absorption data is fitted with $V(t)-V_0$, where $V(t)$ is a time-dependent and $V_0$ is a time-independent Voigt profile (cf. main manuscript). $V_0$ corresponding to each pump laser fluence is obtained using a global fit procedure where a common $V_{0}$ and a time varying $V(t)$ are used to fit several spectra between 200~fs to 1~ps. In the following, the fluence dependence of the $V_{0}$'s parameters are shown and compared with the corresponding magnitude in the steady state absorption of the WS$_2$ monolayer. 

Fig.~\ref{fig:S2} shows the fluence dependence of the different parameters of $V_0$, i.e., the oscillator strength ($S_0$), resonance energy ($E_0$), Lorentzian ($\gamma_0$) and Gaussian ($\sigma_0$) width of both AX and BX. 
The oscillator strength ($S_0$) of both AX (upper panel) and BX (lower panel) increases with pump-fluence but always remains lower than the respective steady state absorption, as indicated by the horizontal line (Fig.~\ref{fig:S2}(a)). The lower magnitude of $S_0$ compared to the steady-state response arises because the photoexcitation of the pump laser beam affects only part of the sample. The fraction increases with increasing excitation density. $S_0$ is expected to approach the steady state value asymptotically with increasing pump laser fluence. The resonance energy, Lorentzian and Gaussian width parameters of $V_0$, respectively (Fig.~\ref{fig:S2}(b-d)), exhibit small variation with pump fluence and are similar to those of the steady state response.  

\section{Fits of the differential absorption spectra}
\begin{figure*}
    \centering
    \includegraphics[width=18cm]{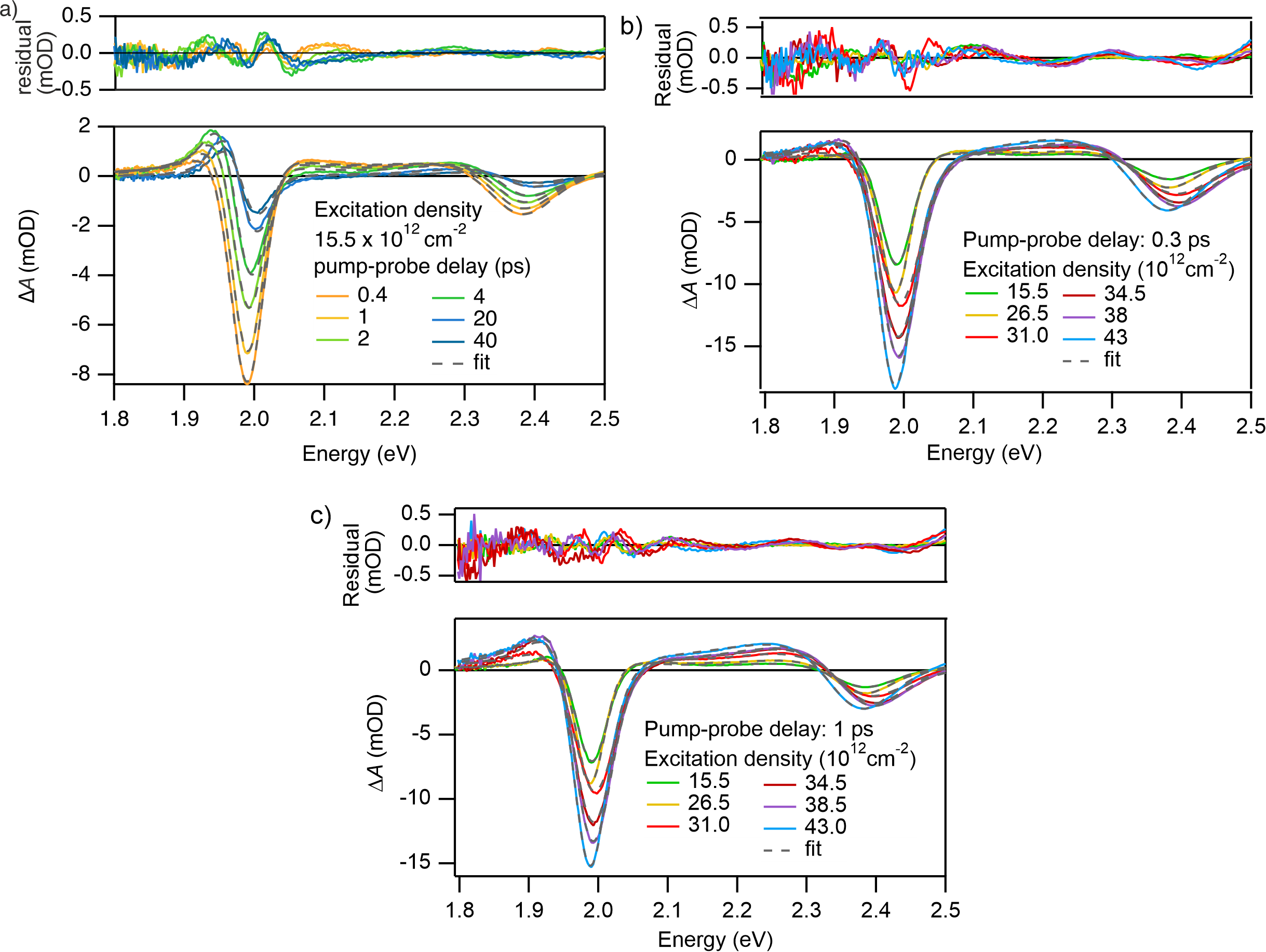}
    \caption{Differential absorption spectra ($\Delta$A) across an energy window capturing the AX and the BX together with fits, i.e. with $V(t)-V_0$, and corresponding residuals. The exemplary traces are shown in (a) for a range of delays (0.4 ps, 1 ps, 2 ps, 4 ps, 20 ps, 40 ps) at an excitation density of 15.5~x~10$^{12}$~cm$^{-2}$ corresponding to a pump fluence of 150 uJ/cm$^2$, (b) for different pump fluence at a delay of 0.3 ps, (c) for different pump fluence at a delay of 1 ps.}
    \label{fig:S3}
\end{figure*}
In Fig.~\ref{fig:S3}(a, b), we show the 
fits of the differential absorption spectra of the WS$_2$ monolayer, using lineshape analysis, for a large range of delays and pump laser fluences, as indicated in the legend. Note that, here and in the rest of the Supplemental Material we use excitation densities instead of pump laser fluences. The conversion between the two quantities as determined in section III.A in the main manuscript is given in Table~\ref{tab:S2}.
\begin{table}[h]
\centering
\small
\setlength{\tabcolsep}{10pt}
\begin{tabular}{@{}cc@{}}
\toprule
\shortstack{Pump laser fluence\\($\mu$J/cm$^2$)} &
\shortstack{Excitation density\\($10^{12}$ cm$^{-2}$)} \\ 
\midrule
47 & 0.75\\ 
152 & 15.5\\ 
292 & 26.5\\ 
461 & 31.0\\ 
653 & 34.5\\ 
768 & 38.0\\ 
1080 & 43.0\\ 
\bottomrule
\end{tabular}
\caption{Conversion from pump laser fluence to excitation density using the band integral analysis as shown in section III. A of the main manuscript.}
\label{tab:S2}
\end{table}

\section{Comparison and fits of differential absorption spectra across the critical excitation density}

\begin{figure*}
    \centering
    \includegraphics[width=10cm]{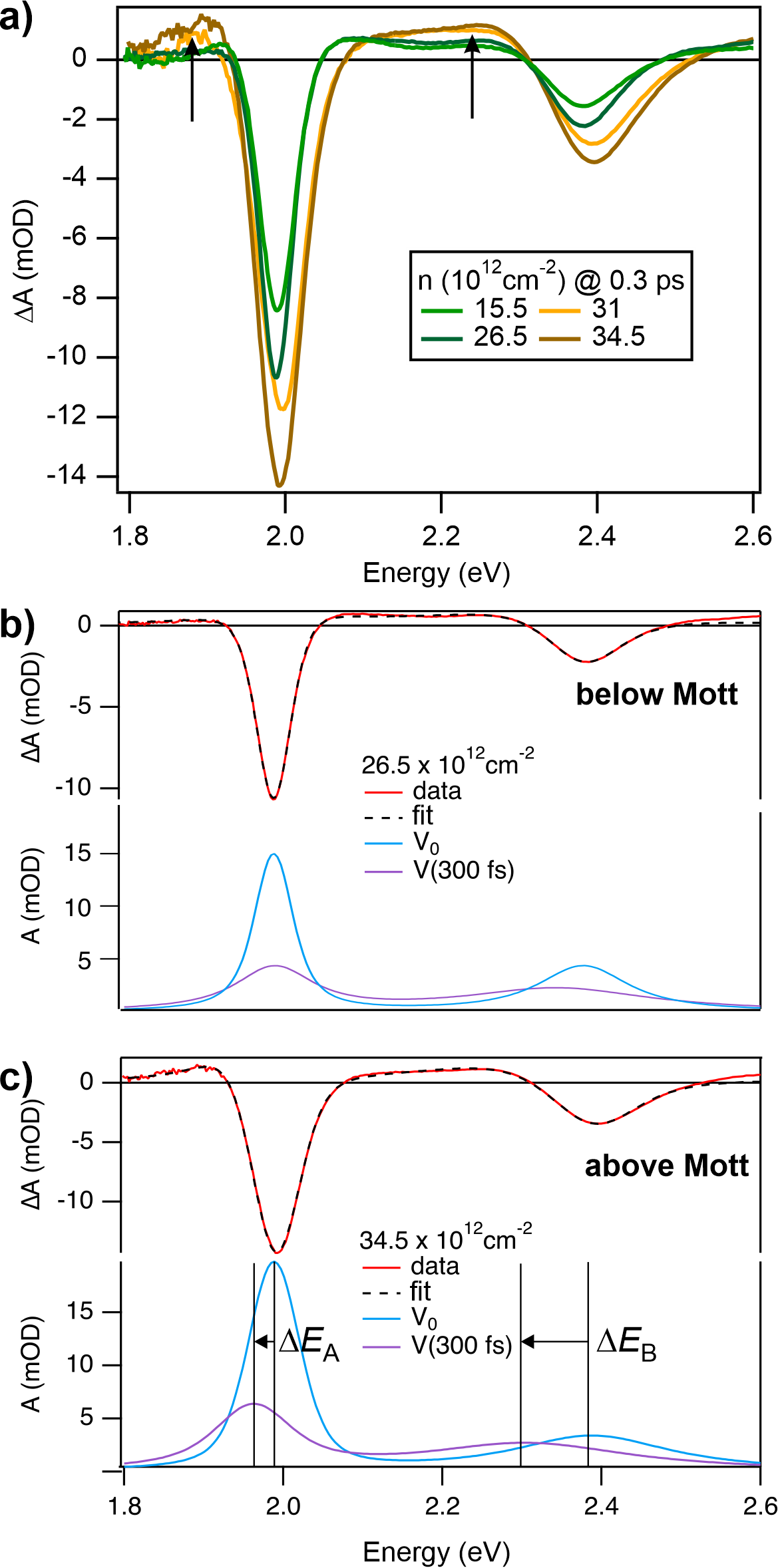}
    \caption{a) Differential absorption spectra ($\Delta$A) below and above the Mott density $n_\text{C}$~=~29~$\pm$~2~x~10$^{12}$~cm$^{-2}$ at a pump probe delay of 0.3~ps. The black arrows highlight the change in the spectra ranges below the AX and BX, attributed to the abrupt red shift of the excitonic resonances. b), c) Corresponding decomposition with the fit model, i.e., $V(t)-V_0$, for two exemplary excitation densities below (panel b) and above (panel c) the Mott density. Above Mott, $V$ shows an increased red-shift with respect to $V_0$ and thereby accounts for the intensity increase below the excitonic resonances. 
    }
    \label{fig:SX}
\end{figure*}

To highlight that the raw data exhibits abrupt changes across the critical density, spectra across the Mott density $n_\text{C}$~=~29~$\pm$~2~x~10$^{12}$~cm$^{-2}$ are compared in the following. Fig. \ref{fig:SX}~a) shows four differential absorption spectra at 0.3~ps out of those presented in the manifold of spectra in Figure \ref{fig:S3}~b). The differential absorption is largest at the AX and BX resonance energies. In the case of the AX at about 2 eV this can be mainly attributed to the ground state bleach, which evolves continuously across the critical density (see section IIIA in the main manuscript). However, at 31.5~x~10$^{12}$~cm$^{-2}$, a clear abrupt increase of intensity below the AX and BX resonance energy is observed (marked with black arrows). Fig. \ref{fig:SX}~b) and c) show the fitting with the lineshape analysis model for two exemplary excitation densities below and above the Mott density. The fit describes the raw data across a broad energy range. Below the Mott density only the BX shows a small red shift with respect to $V_0$, while the AX does not due to the overlapping blue shift resulting from the A exciton population, see section III B 3 in the main manuscript. Above the Mott density, the $V$ contribution at both excitonic resonances is red shifted due to band gap renormalization. This accounts for the intensity increase in the energetic region below both excitonic resonances. The lineshape analysis allows to resolve and track the small but crucial changes in the raw data, which are attributed to the abrupt exciton Mott transition.

\pagebreak

\section{Ground-state bleach at early pump probe delays} 

\begin{figure}
    \centering
    \includegraphics[width=8cm]{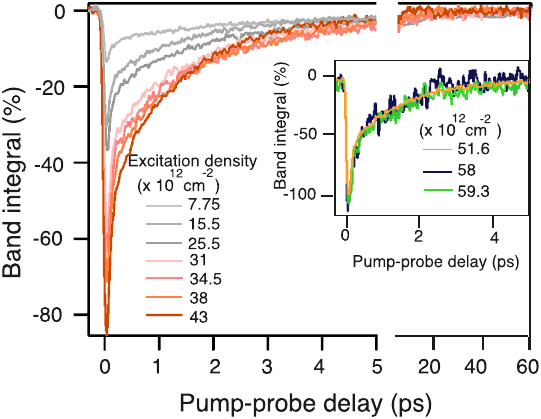}
    \caption{AX band integral kinetics. 
    At high excitation densities a bleach of 100\% is observed close to time zero (see inset).}
    \label{fig:S4}
\end{figure}

In section IIIA of the manuscript, we evaluate the ground state bleach of the AX as a function of pump laser fluence at a fixed delay of 200~fs. To illustrate that the bleach is larger at earlier delays and reaches 100~\% for high fluences we show the temporal evolution of the AX band integral in Fig.~\ref{fig:S4}. For all fluences, the band integral of the AX is reduced with respect to the steady state absorption - the more, the closer to time zero. The maximum bleach is reached at 80~fs. The maximum bleach increases from -20\% to -80\% with increasing the excitation density from 7.75 x 10$^{12}$ cm$^{-2}$ to 43 x 10$^{12}$ cm$^{-2}$. Further increasing the excitation density, the band integral reaches and saturates at 100\% (cf. inset) at this delay. 

\section{Differential absorption spectra at pump-probe overlap}
\begin{figure*}
    \centering
    \includegraphics[width=15cm]{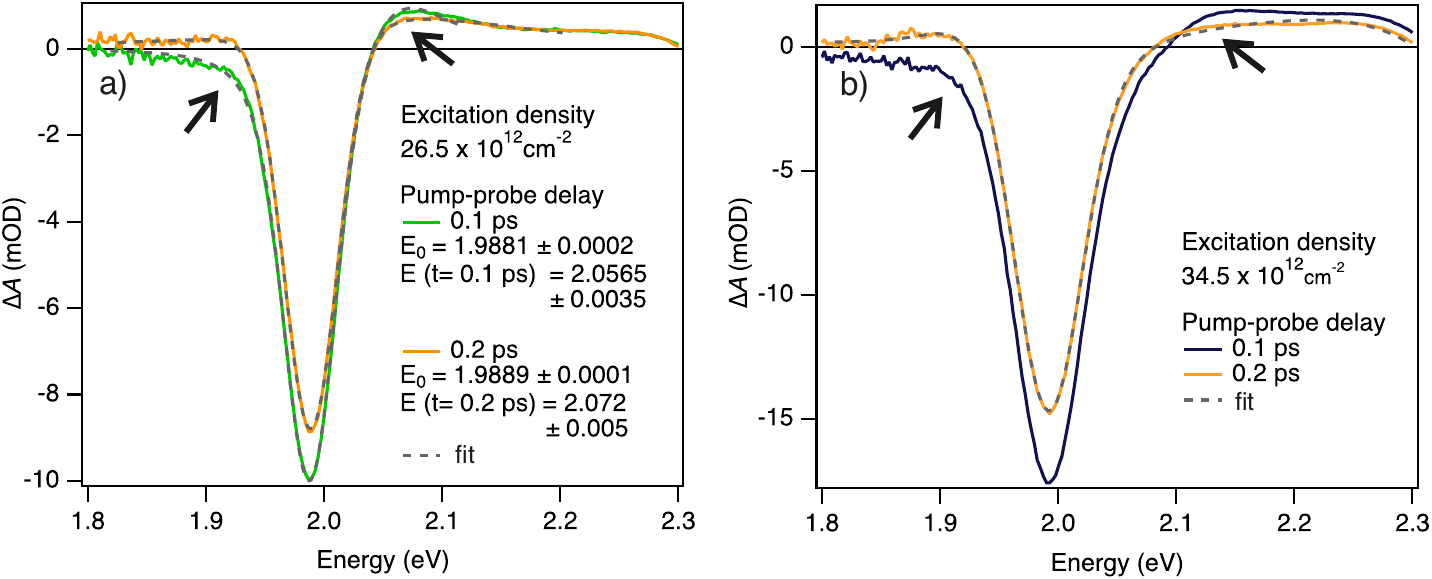}
    \caption{Comparison of the differential absorption ($\Delta$A) spectra at 0.1 ps and 0.2 ps for excitation densities (a) below (26.5 × 10$^{12}$ cm$^{-2}$) and (b) above (34.5 × 10$^{12}$ cm$^{-2}$) the Mott density, along with corresponding fits. The black arrows mark the asymmetry that is visible in particular for early pump-probe delays. At 0.1~ps, the differential absorption is positive at the high energy side and negative on the low energy side for both excitation densities. This is indicative of an overall blue shift in both cases at early delays. The blue shift does not persist up to 0.2~ps in the high excitation density case but apparently decays earlier (see text). }
    \label{fig:S5}
\end{figure*}
As discussed in Section III B3 of the main manuscript, a fast blue shift component decaying with $\tau_\Sigma$=80~fs is observed for the AX in the excitonic regime. Tuning the excitation density just above the Mott density (yellow trace in Fig. 5 (a)), this blue shift component is still observed. However, it is no longer there for even higher excitation densities. The absence of the 80 fs blue shift component could have two explanations: (i) The shift disappears entirely, or (ii) it becomes faster than 80 fs and therefore no longer appears in our analysis, which we restrict in the main text to delays greater than 0.2~ps. To check this, we inspect here the differential absorption spectra closer to time zero. We show that, for low as well as high excitation densities, the spectra at 0.1~ps show the signature of an overall blue shift, thus favoring explanation (ii).

Fig.~\ref{fig:S5}(a) shows the differential absorption spectra for a low excitation density of 26.5 x 10$^{12}$ cm$^{-2}$ at 0.1~ps and 0.2~ps. Both spectra have an asymmetric lineshape: there is a positive feature on the high-energy side, i.e., at $\sim$ 2.1 eV as indicated with the black arrow. At the low energy side, i.e., at $\sim$ 1.9 eV as indicated with another black arrow, the intensity is lower and even negative for 0.1~ps. This is indicative of a blue shift. To crosscheck this qualitative impression, we fit the two TA spectra with $V(t)-V_0$, i.e., with the fitting procedure as described in the main manuscript. The fit is shown with the dashed line in Fig.~\ref{fig:S5}(a). It describes both spectra very well. The AX position is shifted with respect to $V_0$ by 0.069 $\pm$ 0.004 eV and 0.084 $\pm$ 0.005 eV at 0.1 ps and 0.2 ps, respectively. Thus, the fit confirms the visual impression that the AX is blue shifted. 

Fig.~\ref{fig:S5}(b) shows the differential absorption spectra for a high excitation density of 34.5 x 10$^{12}$ cm$^{-2}$ at 0.1~ps and 0.2~ps. The blue-shift related asymmetry of the TA spectrum at 0.1 ps is still visible: there is still a negative signal at lower energies and a positive signal at higher energies. Unfortunately, the spectrum cannot be reliably fitted. However, the clear asymmetry strongly suggests that the blue shift is present at 0.1 ps. By contrast, at 0.2~ps the negative intensity at lower energies has disappeared and so is the pronounced positive signal at higher energies. Fitting with $V(t)-V_0$ indeed does not result in a blue shift, but instead yields a red shift at 0.2 ps (cf. main manuscript Fig. 5(a)). We conclude that the blue shift component decaying with $\tau_\Sigma$ is most likely present for all laser fluences. Above the Mott threshold, it does not appear in our analysis of the spectra at $>$0.2~ps because it is accelerated. 

\section{Peak broadening across the Mott density}

In the main manuscript we focus on the analysis of the AX and BX peak shifts. The broadening also exhibits strong changes at the critical density as discussed in the following. Figure \ref{fig:broadening}~a) and b) shows the broadening $\Delta \gamma_{A,B}$ as a function of pump-probe delay for different excitation densities. In the low fluence regime below the critical excitation density, the trends for AX and BX are monotonic (cf. main manuscript). There is an initial broadening that decays on a timescale of few to tens of ps. Above the Mott density $n_\text{C}$~=~29~$\pm$~2~x~10$^{12}$~cm$^{-2}$, the dynamics change drastically in the first few ps. This can be seen, for example, in the AX broadening at $n_0$~=~43~x~10$^{12}$~cm$^{-2}$ (red trace): the broadening \emph{increases} as a function of pump-probe delay, reaching a maximum at about 1 ps. For the BX, the same trend can be observed for $n_0$~=~31~x~10$^{12}$~cm$^{-2}$ (yellow trace). The timescale of this new dynamics is comparable to the plasma-induced red shift above the Mott density (cf. main manuscript) and is therefore attributed to be induced by the generation of the plasma above $n_\text{C}$. 

To confirm this interpretation, we fit the broadening dynamics with the sum of plasma-, overall population-, and lattice heating- induced broadening, i.e., three exponentials decay functions, analogous to the shift in the main manuscript: 
\begin{figure}
    \centering
    \includegraphics[width=8cm]{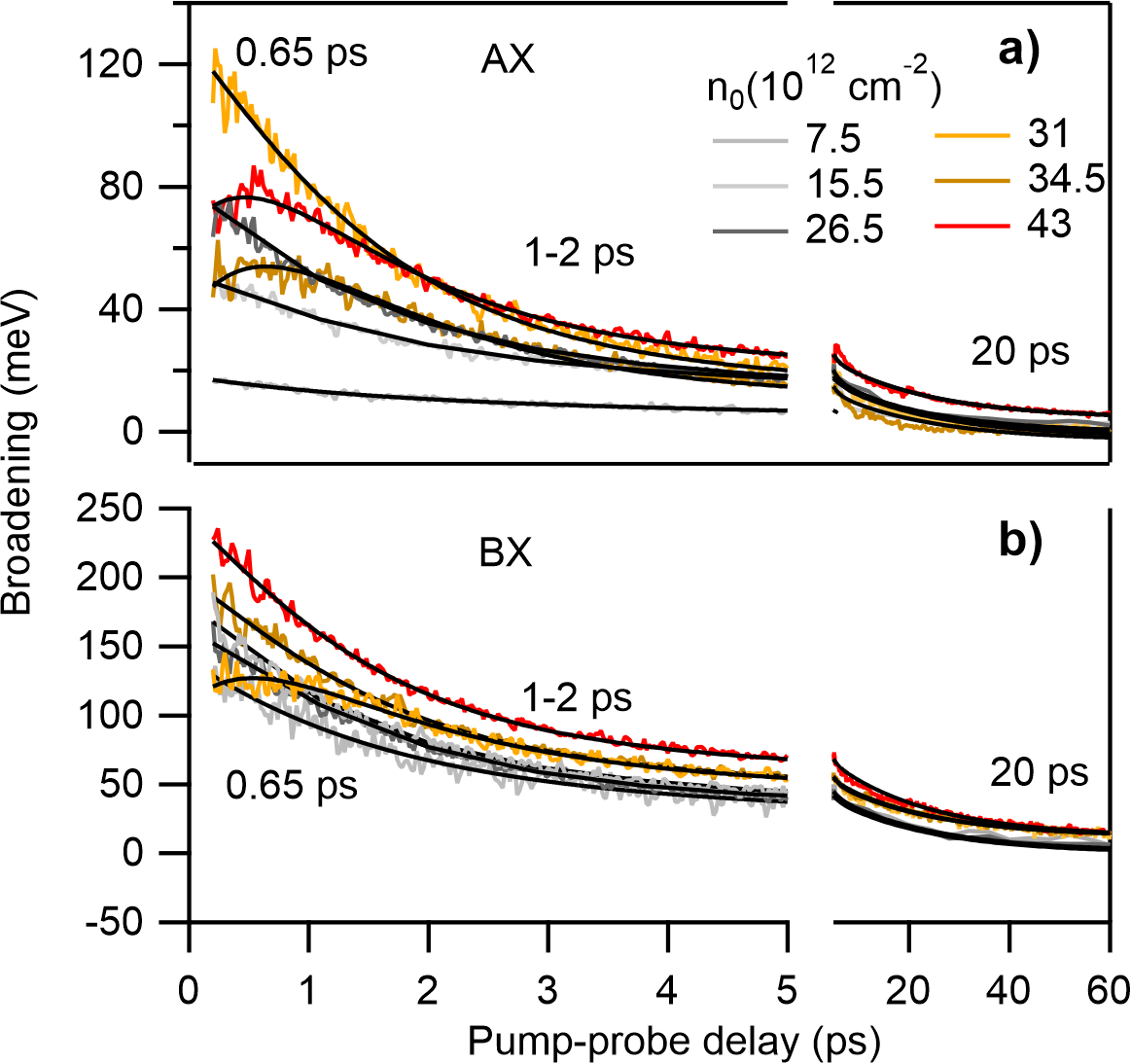}
    \caption{Excitation density-dependence of the peak broadening dynamics. a) and b): Broadening $\Delta \gamma_{A,B}$ for the AX and BX, respectively, as a function of pump-probe delay for different excitation densities as indicated in the legend. Black solid lines are fits with equation \ref{eqn:broadening_fit}.}
    \label{fig:broadening}
\end{figure}

\begin{eqnarray}
\Delta \gamma_{A,B}(t)\!&=&\!D^{A/B}_\text{plasma}\text{e}^{-t/\tau_\text{plasma}} \notag\\
&+&\!D^{A/B}_\text{pop}\text{e}^{-t/\tau_\text{pop}} \notag\\
&+&\!D^{A/B}_\text{heat}\text{e}^{-t/\tau_\text{cool}} + y_0.
\label{eqn:broadening_fit}
\end{eqnarray}

We fix $\tau_\text{plasma}$ to 0.65~ps, the value determined in the main manuscript from analyzing the shifts. Furthermore, we fix $\tau_\text{cool}$ to 20~ps, in agreement with individual fits of shift and broadening shown in the main manuscript. We perform a global fit, where $\tau_\text{pop}$ is globally varied for AX and BX broadening, and at the same time with the area recovery dynamics (main manuscript Fig. 3). The corresponding fits are shown as solid black lines in Figure \ref{fig:broadening}~a) and b) describe the data satisfactorily. This confirms that the broadening can be understood as the combined effect of plasma, overall population, and lattice heating. The plasma decay time accounts for the initial dynamics in the first ps.

\begin{figure}[H]
    \centering
    \includegraphics[width=7cm]{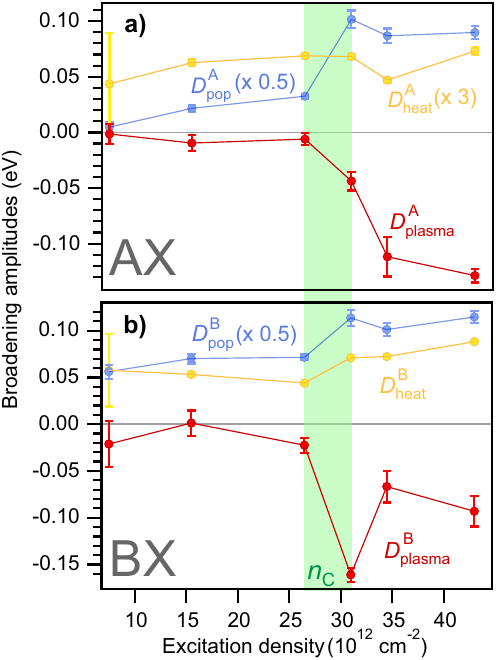}
    \caption{Fit results for the amplitudes of the plasma-induced broadening as a function of excitation density. a) AX, b) BX.}
    \label{fig:broadening_2}
\end{figure}

The fit amplitudes $D_\text{plasma}^\text{A/B}$, $D_\text{pop}^\text{A/B}$ and $D_\text{heat}^\text{A/B}$ are shown for the AX and BX in Figure \ref{fig:broadening_2}~a) and b), respectively. $D_\text{pop}^\text{A/B}$ and $D_\text{heat}^\text{A/B}$ increase with excitation density for the AX and BX. $D_\text{heat}^\text{A/B}$ (yellow) shows no systematic effect at $n_\text{C}$, neither for the AX nor the BX. The trend of $D_\text{pop}^\text{A}$ (blue) insinuates a step-like increase from a finite value before $n_\text{C}$ to a higher value after $n_\text{C}$ for the AX, but this is not as clearly observed for the BX. By contrast, $D_\text{plasma}^\text{A/B}$ (red) shows an important change of behavior: It is zero within error bars below $n_\text{C}$ for both the AX and BX, and it has a large negative value for all excitation densities above $n_\text{C}$. For the BX a prominent dip at $n_\text{C}$ is observed. This illustrates that the plasma contribution becomes significant only at the critical density, in agreement with the analysis of the AX/BX shifts in the main text.

The above demonstrates that at the critical density strong changes occur to the peak broadening of both the AX and BX. The increase of broadening in the time domain means that the plasma leads to less broadening than the excitonic population to which it decays. It is far more difficult to interpret the plasma-induced effect on the peak broadening than on the shift within the EMT. Intuitively, pump-induced free carriers are expected to lead to a stronger broadening of the peaks than photoexcited excitons and not vice versa. Peak broadening is observed, for example, in photoluminescence spectra of MoS$_2$ across the Mott density \cite{bataller2019dense}. However, it should be noted that photoluminescence probes the recombination of plasma carriers, while our measurement is via excitonic resonances dressed in an electron-hole plasma. The interaction between excitons and plasma quasiparticles is expected to involve two competing effects: on the one hand, an increased quasiparticle density and mobility in the plasma phase, and on the other hand, enhanced screening of the Coulomb interaction. Within conventional Fermi-liquid theory, the screening contribution is dominant, resulting in a reduction of electron-electron scattering rates as the plasma density increases \cite{bauer2015}. Although the present case concerns exciton–plasma interactions rather than purely electronic scattering, our observations suggest a similar hierarchy of mechanisms, where the screening induced by the plasma outweighs the effects associated with the increased quasiparticle mobility and density compared to the excitonic phase. In order to properly interpret the effect, additional support from theory would be desirable. Potential changes of scattering coefficients close to the critical density, which have been observed before for a photoinduced EMT \cite{sekiguchi2017anomalous} could play a role. A change of the scattering coefficients as a function of excitation density could e.g. explain that the BX shows a narrowing effect directly at $n_\text{C}$ while for the AX, it becomes prominent only after $n_\text{C}$.




\end{document}